\documentclass[journal=jacsat,manuscript=article,layout=twocolumn]{achemso}
\setkeys{acs}{articletitle=true}
\usepackage{lettrine}

\usepackage[version=3]{mhchem} % Formula subscripts using \ce{}
\usepackage[T1]{fontenc}       % Use modern font encodings

\usepackage[fontsize=10pt]{scrextend}

\usepackage{newtxtext}
\usepackage{bm} 
\usepackage[labelfont=bf]{caption} % bold ``Figure''

\usepackage[hyperfootnotes=false]{hyperref}	 % make refs, cites, links in color
\hypersetup{
  colorlinks,
  citecolor=blue,
  linkcolor=blue,
  urlcolor=blue
}

\usepackage{titlesec} % inline subsections
\titleformat{\subsection}[runin]
{\normalfont\bfseries}{\thesubsection{.}}{1em}{}[.]

\titleformat{\section}
{\normalfont\sffamily\large\bfseries}{\thesection{.}}{1em}{\MakeUppercase}

\usepackage[table]{xcolor}
\usepackage{graphicx}
\usepackage{amsmath,fullpage,amssymb}
\usepackage{graphics,epsfig}
\usepackage{makeidx}
\usepackage{gensymb}
\usepackage{siunitx}

\usepackage{soul}	% for strikethrough font

\def\rme{{\mathrm{e}}}

\def\Eq{eq}
\def\Eqs{eqs}

\newcommand{\Fig}{Figure~\ref}
\newcommand{\Figs}{Figures~\ref}
\newcommand{\suppl}{Supplementary Information}
\newcommand{\methods}{Methods }

\newcommand{\kB}{k_\textrm{B}}

\newcommand{\xx}[1]{\textcolor[rgb]{1,0,0}{[xx]}}

\SectionNumbersOn
\let\oldmaketitle\maketitle
\let\maketitle\relax

\title{Three-phase contact lines as distinct molecular adsorption environments}

\author{Fabio Staniscia}
\affiliation{\rm\small Department of Theoretical Physics, Jo{\v z}ef Stefan Institute, SI-1000 Ljubljana, Slovenia}
\author{Matej Kandu\v{c}}
\affiliation{\rm\small Department of Theoretical Physics, Jo{\v z}ef Stefan Institute, SI-1000 Ljubljana, Slovenia}
\email{matej.kanduc@ijs.si}

\begin{document}
\pagenumbering{arabic}
\noindent

\parindent=0cm
\setlength\arraycolsep{2pt}

\twocolumn[	% make wide abstract
\begin{@twocolumnfalse}
\oldmaketitle

\begin{abstract}\small
In wetting and multiphase systems, three-phase contact lines form where liquid, solid, and gas phases meet, creating effectively one-dimensional boundaries with distinct molecular environments. Despite their ubiquity in droplets, liquid menisci, capillary bridges, porous materials, and confined multiphase systems, molecular adsorption at these lower-dimensional boundaries has remained largely unexplored. 
Here, we use molecular dynamics simulations to show that the contact line on a hydrophobic substrate is not merely the geometric intersection of two interfaces but a distinct thermodynamic adsorption environment with a substantially stronger affinity for amphiphilic molecules than either neighboring interface. 
The investigated amphiphiles of different molecular structures exhibit a common scaling relation between line and surface adsorption in the dilute regime, while their behavior becomes molecule-specific at higher concentrations. Furthermore, contact-line affinity increases exponentially with alkyl-chain length, leading to crowding at progressively lower concentrations and making contact lines extraordinarily sensitive to trace amphiphilic impurities. Because line, surface, and bulk populations scale differently with system size, contact-line adsorption becomes increasingly important at small length scales, where it can substantially alter molecular partitioning. Contact-line enrichment is predicted to reduce the line tension by an amount comparable to its intrinsic magnitude, with potentially substantial consequences for heterogeneous nucleation. Our results establish molecular adsorption as a fundamental property of three-phase contact lines with implications for wetting and multiphase systems across length scales.

%\bf{KEYWORDS:} \sl{Three-phase contact line, Line adsorption, Amphiphiles, Wetting, Line tension, Molecular dynamics simulation}
 %\pacs{...}
\vspace{5ex}
\end{abstract}
\end{@twocolumnfalse}]

\section{Introduction}

Whenever a liquid partially wets a solid in the presence of a gas, a three-phase contact line is formed. Such contact lines are ubiquitous, occurring in droplets, bubbles, porous media, evaporation fronts, capillary bridges, and biological wetting phenomena. Although often viewed simply as the geometric boundary between two interfaces, they constitute lower-dimensional objects with their own excess quantities and thermodynamic relations, analogous to the surface excesses and surface thermodynamics of interfaces~\cite{de1985wetting, drelich1992line, drelich1996significance, schimmele2007conceptual}.

Only recently have three-phase contact lines begun to be recognized as chemically distinct environments in which transport, crystallization, electrochemical processes, nucleation, and chemical reactions become strongly localized~\cite{chen2021three, jimenez2021effect, weng2022engineering, wang2022identification}.
Yet, what determines the molecular affinity for contact lines, how this affinity compares with adsorption at the neighboring interfaces, and how it influences macroscopic phenomena remain poorly understood.

Amphiphilic molecules provide a particularly useful model system with which to address this question. Their adsorption at liquid--vapor, liquid--liquid, and liquid--solid interfaces has been studied extensively for more than a century~\cite{rosen2012surfactants, aveyard2019surfactants}, owing to its central role in colloid science, chemistry, biology, and engineering. In contrast, far less is known about adsorption at three-phase contact lines. 

Resolving this question has important implications. On the one hand, preferential adsorption at the contact line may locally amplify the concentration of specific molecular species and thereby influence reaction pathways, transport processes, or nucleation kinetics.
On the other hand, molecules that preferentially accumulate at one-dimensional interfaces act as line-active agents, also referred to as linactants (also spelled lineactants)~\cite{trabelsi2008linactants}. The concept of lineactants is well established for line boundaries separating coexisting domains in two-dimensional systems (e.g., lipid membranes and Langmuir monolayers)~\cite{schafer2010partitioning, palmieri2014line, bischof2015searching}.  The role of lineactants at three-phase contact lines, however, has received little attention despite occasional theoretical proposals~\cite{law2017line, bey2021carbon}. 
 By adsorbing at the contact line, these molecules contribute directly to the excess free energy of the contact line and are therefore expected to modify the line thermodynamics, analogous to the way surfactants modify surface thermodynamics through interfacial adsorption. Consequently, line adsorption may influence line tension, wetting behavior, contact-angle hysteresis, and nanoscale partitioning. 

To address this knowledge gap, we use molecular dynamics (MD) simulations to investigate the adsorption of amphiphilic molecules at three-phase contact lines.  We show that the contact line provides a distinct and particularly favorable adsorption environment, leading to pronounced molecular enrichment relative to the neighboring interfaces. The adsorption affinity increases exponentially with amphiphile chain length, more strongly than at either interface, which can be explained by a simple continuum picture. Consequently, substantial contact-line adsorption can occur even at trace bulk concentrations, considerably reducing the line tension. We further estimate that the resulting changes in line tension can substantially lower the activation barrier for heterogeneous bubble nucleation, with potentially significant consequences for cavitation and bubble formation.

\section{Results}
\subsection{Surface and line adsorption}
\begin{figure*}
  \includegraphics[width=0.99\linewidth]{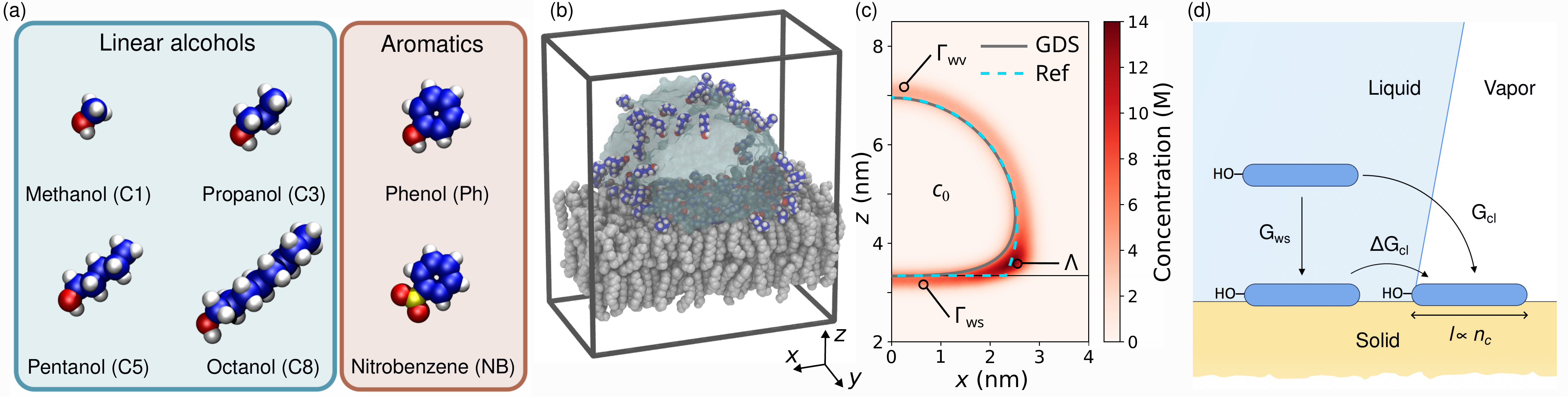}  
  \caption{\textbf{Preferential accumulation of amphiphiles at the three-phase contact line.} (a) Amphiphilic molecules investigated in this study, grouped into four linear amphiphiles (alcohols) and two aromatic compounds.
(b) Snapshot of the simulation setup showing a cylindrical-cap water nanodroplet on a solid substrate with 96 dissolved C5 molecules. Water is represented by a transparent smoothed density isosurface.
(c) Cross-sectional concentration map of C5 at a bulk concentration of $c_0 \approx 5$ mM. C5 accumulates at both the water--vapor ($\Gamma_\mathrm{wv}$) and water--solid ($\Gamma_\mathrm{ws}$) interfaces, with particularly strong enrichment at the three-phase contact line ($\Lambda$). The horizontal solid line marks the effective substrate boundary, the solid gray curve denotes the Gibbs dividing surface (GDS) of the water phase, and the cyan dashed line defines the reference droplet used in the evaluation of \Eq~\ref{eq:line_ads}. 
Details of the analysis are given in the \methods section. 
(d) Schematic illustration of amphiphile adsorption at the three-phase contact line. Adsorption from the bulk to the water--solid interface involves a free-energy change $G_\mathrm{ws}\propto -n_\mathrm{C}$, where $n_\mathrm{C}$ is the alkyl-chain length. Subsequent transfer to the contact line, where the hydrophobic chain protrudes into the vapor phase, produces an additional free-energy change $\Delta G_\mathrm{cl}\propto -n_\mathrm{C}$. The total adsorption free energy at the contact line is therefore $G_\mathrm{cl}=G_\mathrm{ws}+\Delta G_\mathrm{cl}$. 
}
 \label{fig:densities}
\end{figure*}

We first ask whether the three-phase contact line provides an adsorption environment for dissolved molecules distinct from the adjacent interfaces. Here, we performed molecular dynamics simulations of a homologous series of amphiphilic molecules, shown in \Fig{fig:densities}a. The series comprises the linear alcohols methanol (C1), propanol (C3), pentanol (C5), and octanol (C8), together with the aromatic molecules phenol (Ph) and nitrobenzene (NB). 
The alcohols allow us to vary the hydrophobic part systematically while keeping the strongly hydrated hydroxyl headgroup unchanged, thereby isolating the effect of the hydrophobic moiety, which primarily governs interfacial partitioning.
The two aromatic molecules allow us to examine whether the observed trends also apply beyond the homologous series, with nitrobenzene, a less conventional amphiphile, providing a particularly interesting test of their generality.

To investigate amphiphile adsorption at a three-phase contact line, we consider a liquid wedge formed where water, vapor, and a solid substrate meet. We realize this geometry using a cylindrical-cap water nanodroplet on an atomistically smooth hydrophobic substrate with a contact angle of $\theta\approx120^\circ$ (\Fig{fig:densities}b). The droplet extends along the cylinder axis, forming two straight, parallel contact lines. This geometry allows us to quantify adsorption per unit contact-line length while avoiding the additional effects of contact-line curvature. 
Although curvature may influence adsorption for nanoscopic radii, the straight-line geometry provides a reference for the limit of vanishing contact-line curvature.
 
The concentration map for C5 in \Fig{fig:densities}c shows that amphiphiles adsorb at both the water--vapor and water--solid interfaces. More strikingly, they accumulate even more strongly at the contact line, where the local concentration substantially exceeds that at either neighboring interface. Such preferential accumulation is the defining characteristic of line-active molecules (lineactants). This behavior is observed for all amphiphiles investigated, including NB. Analysis of the molecular orientations (see Section S1 in the \suppl~(SI)) further reveals that, except for NB, the polar headgroups remain hydrated within the droplet, whereas the hydrophobic moieties extend onto the substrate approximately perpendicular to the contact line, as schematically illustrated in \Fig{fig:densities}d.

Before quantifying adsorption at the contact line, we characterize adsorption at the two neighboring interfaces. To do that, we partition the droplet into bulk, interfacial, and external regions, as described in the Methods section. The surface excesses at the water--vapor ($\Gamma_\mathrm{wv}$) and water--solid ($\Gamma_\mathrm{ws}$) interfaces are then obtained from the amphiphile concentration profiles using the generalized Gibbs dividing surface for curved interfaces (shown by a solid curve in \Fig{fig:densities}c)~\cite{Staniscia25}. 

\begin{figure*}[t]
    \includegraphics[width=0.99\linewidth]{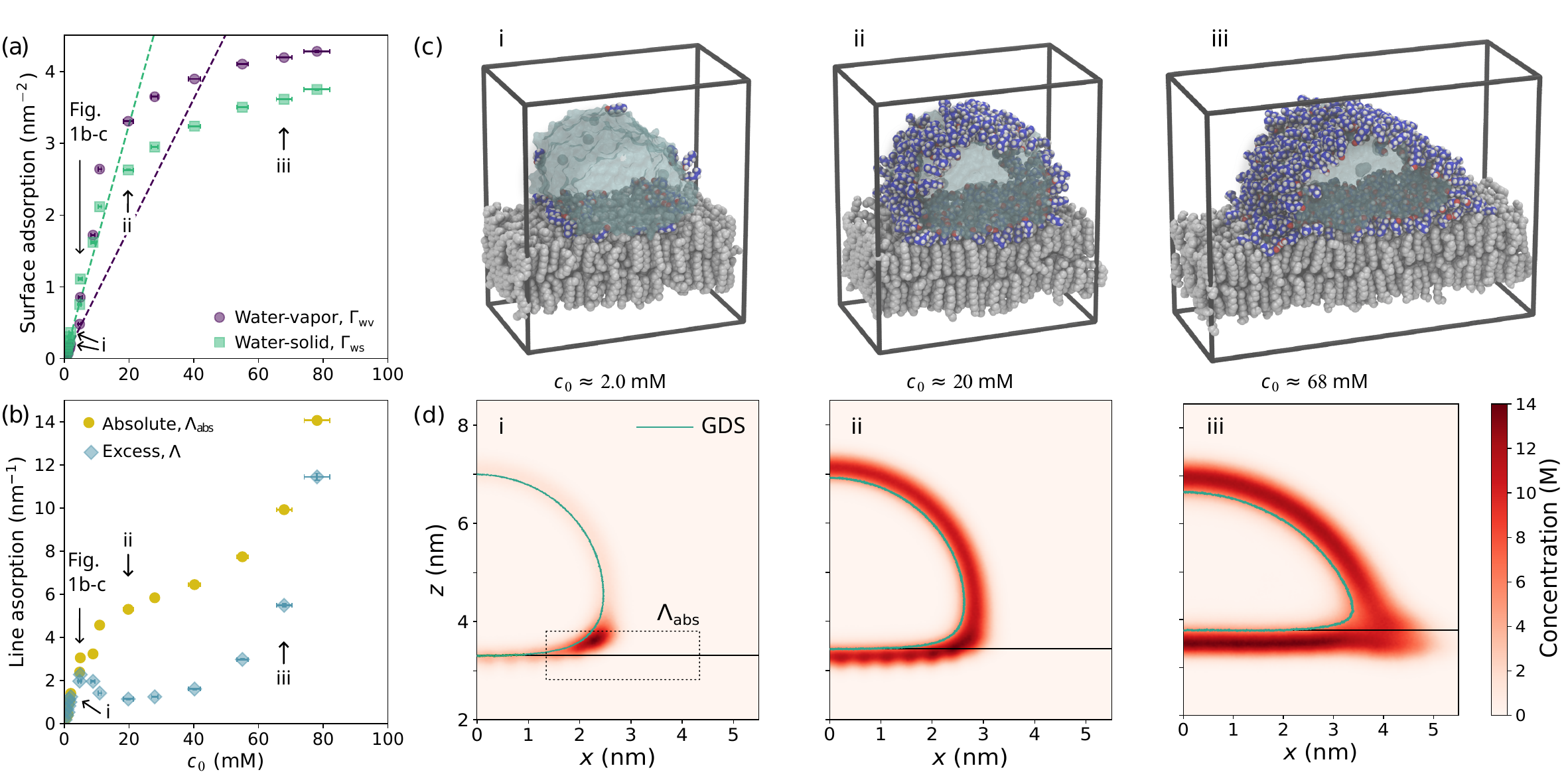}\\ 
  \caption{
  \textbf{Surface and contact-line adsorption of amphiphiles.}
  (a) Surface adsorption isotherms for C5 at the water--vapor and water--solid interfaces (symbols). Dashed lines show fits to the low-concentration Henry regime (eq~\ref{eq:Gamma_linear}).
(b) Absolute line adsorption, $\Lambda_{\mathrm{abs}}$, and (excess) line adsorption, $\Lambda$, defined according to eq~\ref{eq:line_ads}, as a function of bulk concentration. 
(c) Simulation snapshots and (d) corresponding concentration maps at the concentrations labeled i--iii in panels (a) and (b). The concentration corresponding to \Figs{fig:densities}b and \ref{fig:densities}c is also indicated.  The dotted rectangle ($1~\mathrm{nm}\times 3~\mathrm{nm}$) in panel (d)i  indicates the region in which molecules are counted to calculate the absolute line adsorption $\Lambda_{\mathrm{abs}}$.}
  \label{fig:alc_line_ads}
\end{figure*}

The resulting adsorption isotherms, shown for C5 in \Fig{fig:alc_line_ads}a (see Section S2 in the SI for the remaining amphiphiles), exhibit a linear dependence on the bulk concentration $c_0$ at low concentrations, consistent with Henry's law,
\begin{equation}
    \label{eq:Gamma_linear}
    \Gamma_\alpha = K_\alpha\,c_0
\end{equation}
where $K_\alpha$ is the \emph{surface adsorption coefficient} of interface $\alpha$, quantifying the affinity of an individual amphiphile molecule for that interface.

At higher concentrations, the adsorption deviates from Henry's law and gradually approaches a plateau corresponding to monolayer saturation. For the alcohols, the saturation coverage is $\Gamma_\infty\approx3$--$4~\mathrm{nm}^{-2}$, consistent with theoretical and experimental studies of surfactant adsorption~\cite{Staniscia22a, Bleys1985}. This value follows naturally from molecular packing: for a characteristic molecular size of $\sigma\approx0.5$ nm, one expects $\Gamma_\infty\sim1/\sigma^2$. 
Representative snapshots and concentration maps are provided in \Figs{fig:alc_line_ads}c and \ref{fig:alc_line_ads}d, showing a gradual increase of surface excess with concentration.

We now turn to adsorption at the contact line. As a first, intuitive measure, we define the \emph{absolute line adsorption}, $\Lambda_{\mathrm{abs}}$, as the number of amphiphiles contained within a rectangular slab surrounding the contact line, $N_\mathrm{in}$ (shown in \Fig{fig:alc_line_ads}d(i) by a dotted rectangle),  divided by the contact-line length (i.e., $\Lambda_\mathrm{abs}=N_\mathrm{in}/L_y$, for a single contact line).
The resulting absolute adsorption isotherm is shown in \Fig{fig:alc_line_ads}b (circle symbols) and increases monotonically with bulk concentration. 

Like surface adsorption, $\Lambda_{\mathrm{abs}}$ increases linearly in the dilute regime.
Notably, the contact-line region becomes crowded well before either interface, as seen in the representative snapshot in \Fig{fig:alc_line_ads}c(i). Consequently, $\Lambda_\mathrm{abs}$ departs from its initial linear behavior at substantially lower bulk concentrations than the corresponding surface isotherms (cf.\ \Fig{fig:alc_line_ads}a), meaning that amphiphiles preferentially populate the contact line first. Once the contact-line region becomes crowded, the increase in $\Lambda_\mathrm{abs}$ with $c_0$ slows down. As the neighboring interfaces subsequently approach saturation, the relative contribution of contact-line adsorption increases, leading to a renewed increase in the slope of the isotherm. At the highest concentrations, the slope rises sharply, exhibiting an apparent singular behavior as the solubility limit is approached. The corresponding concentration map (\Fig{fig:alc_line_ads}d(iii)) reveals the formation of a growing amphiphile-rich foot extending from the contact line, reminiscent of a precursor film.

Although intuitive, $\Lambda_{\mathrm{abs}}$ is not a thermodynamic excess quantity, since its value depends on the arbitrary choice of the integration region at the contact line. We therefore introduce the \emph{line adsorption}, $\Lambda$, defined as the excess number of amphiphiles per unit contact-line length, in direct analogy to the Gibbs surface excess, $\Gamma$.

To define $\Lambda$, we adopt the water Gibbs dividing surfaces (GDS) for both interfaces and construct an ideal cylindrical-cap reference geometry (dashed cyan curve in \Fig{fig:densities}c). This geometry is obtained by fitting a circle to the water--vapor GDS and extending it to the vapor--solid interface. Unlike the rounded contact-line region observed in MD, the reference geometry has a sharp contact line and defines the reference volume and interfacial areas.
The total number of amphiphiles, $N$, can then be decomposed into bulk, surface, and line contributions,
\begin{equation}
\label{eq:line_ads}
c_0 V + \Gamma_\mathrm{wv} A_\mathrm{wv} + \Gamma_\mathrm{ws} A_\mathrm{ws} + \Lambda l_{\mathrm{cl}} = N .
\end{equation}
Here, $V$ is the reference droplet volume,  $A_\mathrm{wv}$ and $A_\mathrm{ws}$ are the water--vapor and water--solid interfacial areas, and $l_\mathrm{cl}$ is the contact-line length. The first three terms represent the bulk and surface contributions, while the remaining term defines the (excess) line adsorption, $\Lambda$. The cylindrical droplets considered here have two equivalent straight contact lines, therefore $l_\mathrm{cl}=2L_y$.

Unlike $\Lambda_\mathrm{abs}$, $\Lambda$ defined by \Eq~\ref{eq:line_ads} is independent of the arbitrary choice of the integration region surrounding the contact line. However, as with the Gibbs surface excess, its value depends on the choice of dividing surfaces,  which are consistently fixed here by the equimolar convention for water.

\begin{figure*}[h]
    \includegraphics[width=0.34\linewidth]{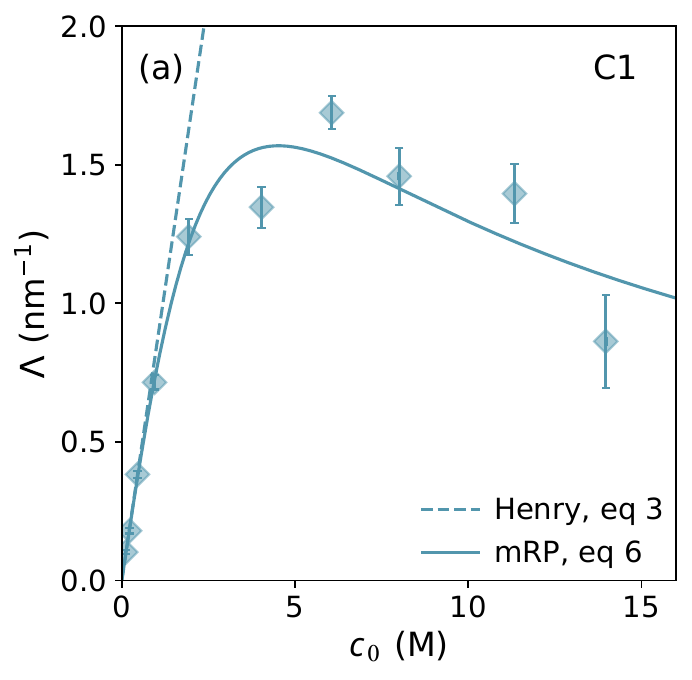}
    \includegraphics[width=0.306\linewidth]{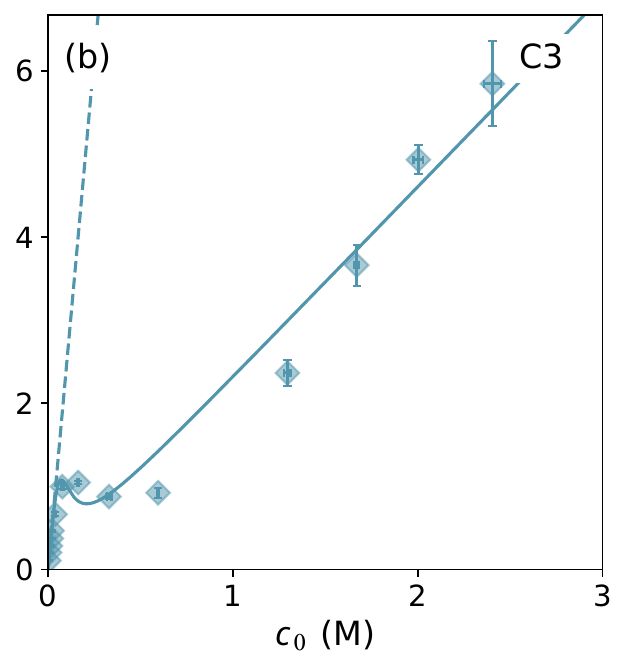}
        \includegraphics[width=0.301\linewidth]{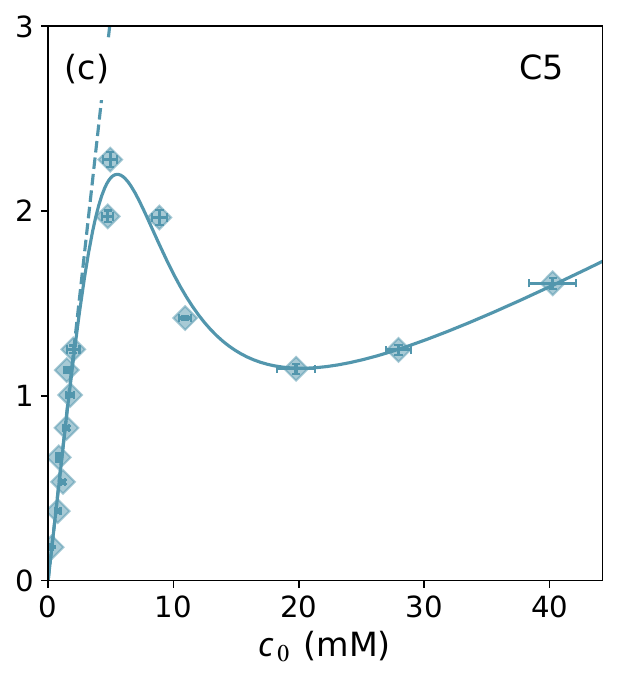}\\
\includegraphics[width=0.336\linewidth]{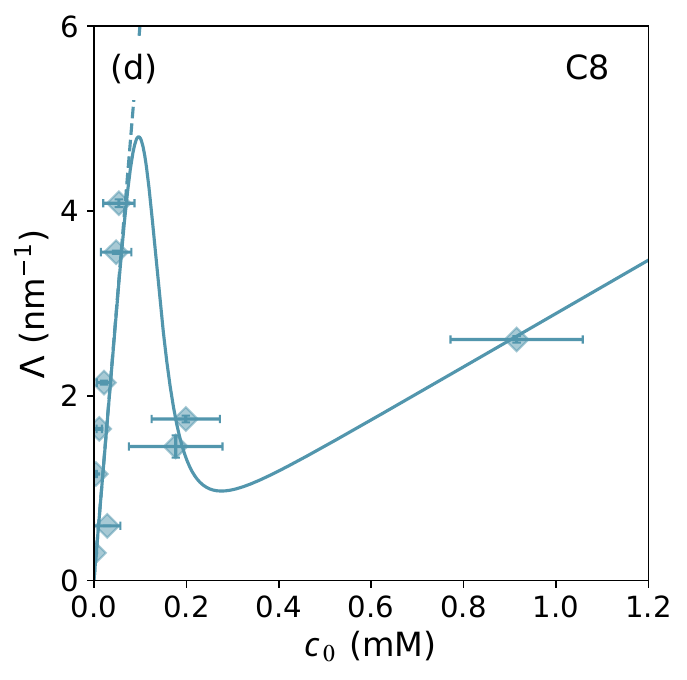}
\includegraphics[width=0.320\linewidth]{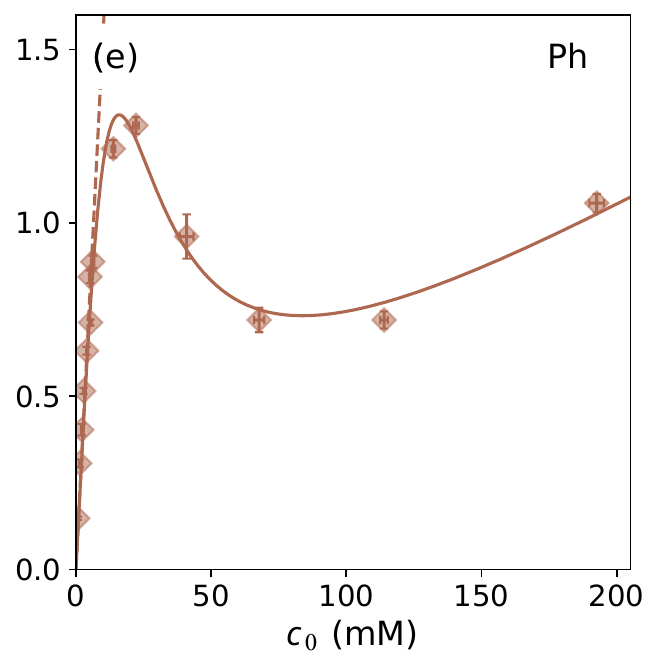}
  \includegraphics[width=0.298\linewidth]{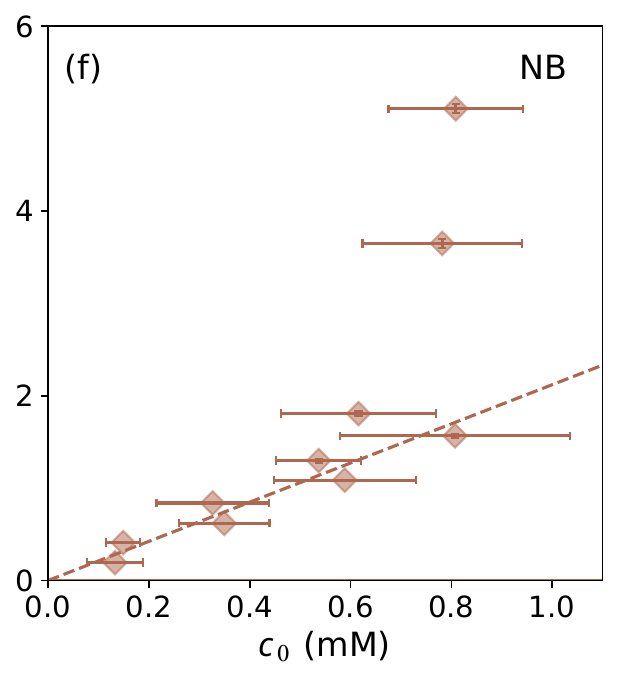}
  \caption{\textbf{Line adsorption isotherms reveal the onset of contact-line crowding.} Line adsorption isotherms for (a) C1, (b) C3, (c) C5, (d) C8, (e) Ph, and (f) NB (symbols). Dashed lines show fits to the low-concentration Henry regime (\Eq~\ref{eq:henry_line}), while solid lines show fits of the mRP isotherm (\Eq~\ref{eq:langmiur_line}) over the entire concentration range (NB is excluded). 
}
  \label{fig:other_alc_line_ads}
\end{figure*}

The resulting line adsorption isotherm, $\Lambda(c_0)$, is shown in \Fig{fig:alc_line_ads}b (diamond symbols) and exhibits a nonmonotonic behavior.
In the dilute regime, $\Lambda$ increases linearly with concentration, closely following $\Lambda_\mathrm{abs}$. The initial increase is followed by a pronounced maximum, a broad minimum, and finally a gradual increase at higher concentrations.

This nonmonotonic behavior stems from the competition between adsorption at the contact line and at the neighboring interfaces.  At intermediate concentrations, the absolute line adsorption enters a regime of slower growth (\Fig{fig:alc_line_ads}b), whereas adsorption at both interfaces continues to increase approximately linearly (\Fig{fig:alc_line_ads}a). As a result, the excess associated with the contact line decreases, producing the local maximum and subsequent minimum in $\Lambda$. As the interfaces gradually approach saturation, however, the contact line again becomes the preferred adsorption site, causing the excess line adsorption to increase.
At the highest concentrations, $\Lambda$ exhibits the same steep upturn observed for $\Lambda_\mathrm{abs}$, coinciding with the development of a precursor-film-like foot.
Since this high-concentration regime lies close to the solubility limit and is beyond the scope of the present work, we restrict the following analysis to concentrations below the onset of this secondary increase.

Line adsorption for all amphiphiles (including C5) up to the intermediate concentrations is shown in \Fig{fig:other_alc_line_ads}. 
For the homologous alcohol series, the shape of the isotherms evolves systematically with increasing chain length. C1 and C3 exhibit only a shallow maximum, whereas for C5 and C8 the peak becomes progressively more pronounced, followed by a minimum and a subsequent rise at higher concentrations. At the same time, the local maximum shifts systematically toward lower concentrations. 

A similar nonmonotonic behavior is observed for phenol, suggesting that this adsorption pattern extends beyond linear alcohols to other surfactant-like molecules. Nitrobenzene, however, provides an instructive counterexample. Unlike the other amphiphiles, it shows no well-defined preferred orientation at the contact line (Section S1 in the SI) and exhibits a much stronger affinity for the water--solid than for the water--vapor interface (Section S2). At $c_0 \simeq 0.8$ mM, its water--solid adsorption increases sharply, accompanied by a rapid increase in the line adsorption $\Lambda$, without the characteristic intermediate maximum. 

Thus, while all investigated molecules share a common adsorption scaling in the dilute regime, their behavior at higher concentrations is not universal. The similarity between alcohols and phenol, which exhibit a preferential amphiphilic orientation at the contact line, suggests a broader trend among surfactant-like molecules. In contrast, nitrobenzene does not show such preferential orientation, consistent with its less pronounced amphiphilic character and distinct adsorption behavior at higher concentrations.

\subsection{Dilute contact-line adsorption}
At low concentrations, interactions between adsorbed amphiphiles are negligible, and the line adsorption follows Henry's law,
    \begin{equation}
        \label{eq:henry_line}
        \Lambda = L\, c_0   \, ,
    \end{equation}
where $L$ is the {\it line adsorption coefficient},  the one-dimensional analogue of the surface adsorption coefficient (\Eq~\ref{eq:Gamma_linear}).
It quantifies the affinity of an individual amphiphile molecule for the bare contact line, and its values, obtained from fits to the low-concentration regime in \Fig{fig:other_alc_line_ads}, are listed in Table~\ref{tab:parameters}.

\begin{table*}[htbp]
\centering
\footnotesize
\caption{Parameters for individual amphiphiles extracted from simulation data. The crossover line tension reduction, $\Delta \tau^*$, is calculated using eq~\ref{eq:dtau*}.}
%\resizebox{0.8\textwidth}{!}{%
\begin{tabular}{lccccccccc}
\hline
\rule{0pt}{2.5ex}& $K_\mathrm{wv}$ (nm)  & $K_\mathrm{ws}$ (nm) & $L$ (nm$^2$)  & $L_1$ (nm$^2$)  & $L_2$ (nm$^2$) & $c^*$ (M) &  $\beta$  & $\Delta \tau^*$ (pN)\\
 \hline
Methanol (C1)& $0.80 (2) $ & $1.18(3)  $ & $1.40 (5)$ &   --- & --- & $3.7 (1)$ &  $1.7 $ & $-25$ \\
Propanol (C3) & $17.4 (4)$ &  $21.5 (5)$ & $41 (1)$  & $37 (1)$ & $3.8 (2)$ & $7.9 (3)  \times 10^{-2}$ &  $2.7 $ & $-10$ \\
Pentanol (C5) & $1.5 (1) \times 10^2$  & $2.7 (2) \times  10^2$ & $1.01 (9) \times  10^3$ & $9.5  (9)  \times  10^2$ & $62(3)  $  & $6.6 (1) \times 10^{-3}$ &  $2.9 $  & $-20$ \\
Octanol (C8) & $3(2) \times10^3$ & $1.1(7) \times10^4$ & $1.0(7) \times10^5$ & $9(7) \times10^4$ & $4.8(7) \times10^3$ & $1.3(2) \times10^{-4}$   &  $5.7$ &$-33$\\
Phenol (Ph) & $37 (2)$ & $1.52 (8) \times 10^2$ & $2.6 (2) \times 10^2$  & $2.5 (2) \times 10^2$  & $7.7 (4)$ &  $1.65  (3)\times 10^{-2}$  & $2.2 $ & $-15$\\
Nitrobenzene (NB) & $77 (9)$ & $1.1(1) \times 10^3$ & $3.5(4) \times 10^3$ & --- & --- & --- & --- & --- \\
\hline
\end{tabular}
%}
\label{tab:parameters}
\end{table*}

\begin{figure*}[h]
  \includegraphics[width=0.33\linewidth]{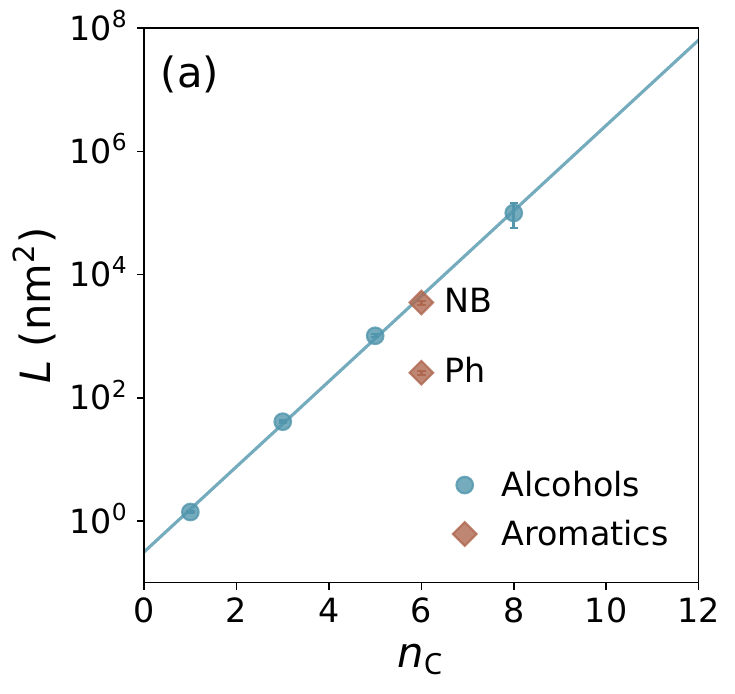}
   \includegraphics[width=0.317\linewidth]{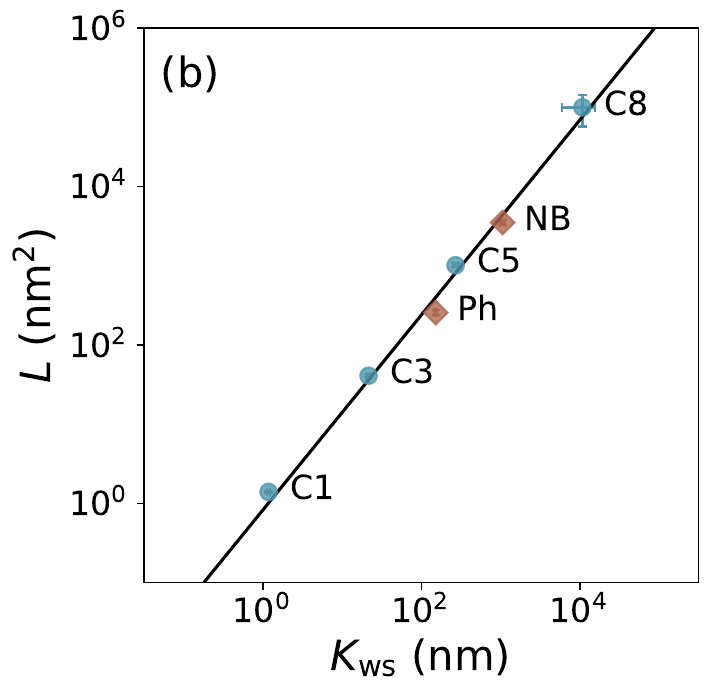}
     \includegraphics[width=0.335\linewidth]{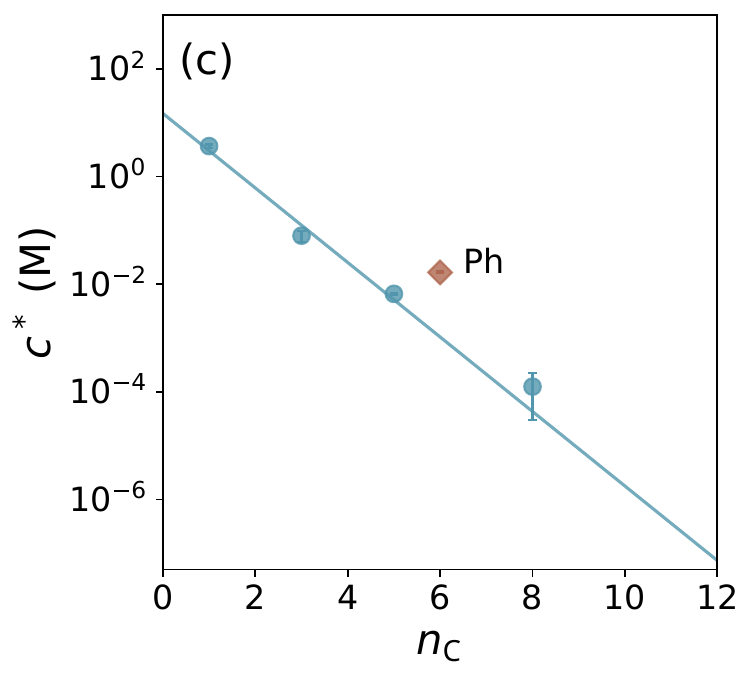}
  \caption{
  \textbf{Scaling relations in contact-line adsorption.}
(a) Line adsorption coefficient as a function of the number of carbon atoms, $n_\mathrm{C}$. The solid line shows a fit of eq~\ref{eq:exp_lin_ads} to the  alcohols.
(b) Line adsorption coefficient as a function of the water--solid surface adsorption coefficient, $K_\mathrm{ws}$. The solid line shows the scaling relation from eq~\ref{eq:scaling3}. (c) Crossover concentration obtained from fits of the mRP isotherm (\Eq~\ref{eq:langmiur_line}) to the line adsorption isotherms, as a function of the number of carbon atoms. The solid line shows the exponential scaling predicted by eq~\ref{eq:c*}, with only $c^{*(0)}$ as the fitting parameter.}
\label{fig:KL_vs_nc}%  \label{fig:cstar}
\end{figure*}

The line adsorption coefficient depends strongly on amphiphile hydrophobicity. As shown in \Fig{fig:KL_vs_nc}a for linear amphiphiles, $L$ increases exponentially with the number of carbon atoms in the alkyl chain, $n_\mathrm{C}$, following
\begin{equation}
\label{eq:exp_lin_ads}
    L = L^{(0)} \, \mathrm{e}^{\alpha_\mathrm{cl} n_\mathrm{C}} \, ,
\end{equation}
with $\alpha_\mathrm{cl} \approx 1.59$ obtained from the fit (solid line). This corresponds to an approximately fivefold increase in line adsorption affinity for each additional CH$_2$ group.

A similar exponential scaling holds for surface adsorption, with $K_\mathrm{wv} \simeq K_\mathrm{wv}^{(0)} \mathrm{e}^{\alpha_\mathrm{wv} n_\mathrm{C}}$ and $K_\mathrm{ws} \simeq K_\mathrm{ws}^{(0)} \mathrm{e}^{\alpha_\mathrm{ws} n_\mathrm{C}}$~\cite{Staniscia22a, Staniscia25}. From the present simulations (Figure S5a in the SI), we obtain $\alpha_\mathrm{wv}\approx1.17$ and $\alpha_\mathrm{ws}\approx1.29$, corresponding to an approximately threefold increase in surface adsorption affinity per additional CH$_2$ group, consistent with the classical Traube's rule. Importantly, $\alpha_\mathrm{cl}>\alpha_\mathrm{ws},\alpha_\mathrm{wv}$, demonstrating that increasing hydrophobic chain length enhances adsorption at the contact line more strongly than at either neighboring interface.

These exponential scaling relations can be understood from simple continuum arguments based on the molecular arrangement illustrated in \Fig{fig:densities}d. Adsorption at the water--solid interface reduces the unfavorable contact between the hydrophobic chain and water, yielding an adsorption free energy that becomes more negative approximately linearly with chain length, $G_\mathrm{ws}\propto -n_\mathrm{C}$~\cite{ramezani2026hydration}. At the contact line, the hydrophobic chain is further exposed to the vapor phase, providing an additional free-energy gain, $\Delta G_\mathrm{cl}\propto -n_\mathrm{C}$. The total adsorption free energy is therefore $G_\mathrm{cl}=G_\mathrm{ws}+\Delta G_\mathrm{cl}$, which decreases more rapidly with chain length than $G_\mathrm{ws}$. In the dilute regime, the adsorption coefficients follow Boltzmann scaling, $K_\mathrm{ws}\propto\exp(-G_\mathrm{ws}/k_\mathrm{B}T)$ and $L\propto\exp(-G_\mathrm{cl}/k_\mathrm{B}T)$, explaining why $\alpha_\mathrm{cl}>\alpha_\mathrm{ws}$. A detailed continuum model presented in Section S3 in the SI provides a semiquantitative estimate of the difference between these exponents, consistent with the MD results.

Although phenol (Ph) and nitrobenzene (NB) do not belong to the homologous alcohol series, they closely follow the same chain-length trend. More importantly, when the line adsorption coefficient, $L$, is plotted against the water--solid surface adsorption coefficient, $K_\mathrm{ws}$, all amphiphiles collapse onto a single master curve (\Fig{fig:KL_vs_nc}b). 

For alcohols, this relation follows directly from the exponential scaling of the two adsorption coefficients.
Eliminating $n_\mathrm{C}$ from \Eq~\ref{eq:exp_lin_ads} and the expression for $K_\mathrm{ws}$ yields
\begin{equation}
\label{eq:scaling3}
    \frac{L}{L^{(0)}}=\left(\frac{K_\mathrm{ws}}{K_\mathrm{ws}^{(0)}}\right)^{\alpha_\mathrm{cl}/\alpha_\mathrm{ws} }\, .
\end{equation}
The prediction, $L \propto K_\mathrm{ws}^{1.23}$ (solid line in \Fig{fig:KL_vs_nc}b), agrees remarkably well with the simulation data. Even nitrobenzene, which exhibits distinctly different adsorption behavior at higher concentrations, follows the same scaling relation in the dilute regime. By comparison, the correlation with the water--vapor adsorption coefficient, $K_\mathrm{wv}$, is somewhat weaker (see Figure~S5b in the SI).

These results indicate that the affinity of amphiphiles for the contact line  on a hydrophobic substrate is governed primarily by their affinity for the water--solid interface, despite their chemical diversity. Consequently, line adsorption can be estimated from the experimentally more accessible surface adsorption coefficient. The continuum model in Section S3 in the SI provides a physical basis for this scaling and predicts that the exponent $\alpha_\mathrm{cl}/\alpha_\mathrm{ws}$ depends on substrate wettability.

Although our main analysis focuses on a strongly hydrophobic substrate, additional simulations for C3 on substrates characterized by lower contact angles $\theta$, in Section S4 of the SI, show that both line and water--solid adsorption decrease as the substrate becomes more hydrophilic. Specifically, both adsorption coefficients, $L$ and $K_\mathrm{ws}$, vary approximately exponentially with $\cos\theta$, as predicted by our continuum model (Section S3 in the SI), which yields adsorption free energies that are linear in $\cos\theta$. For even more hydrophilic substrates, amphiphiles may spread out of the droplet, forming a precursor-like film.

\subsection{Contact-line crowding}

As the concentration of amphiphiles increases beyond the dilute regime, the contact line gradually becomes crowded, and the adsorption isotherms deviate from Henry's law.  To describe the adsorption behavior beyond the dilute regime, we introduce a simple phenomenological expression,
\begin{equation}\label{eq:langmiur_line}
    \Lambda =  \frac{L_1 \, c_0}{1 + (c_0/c^*)^{\beta}} + L_2\, c_0
\end{equation} 
which we refer to as the modified Redlich--Peterson (mRP) isotherm.
The first term has the form of the classical Redlich--Peterson isotherm and captures the initial increase in line excess, the maximum, and subsequent decrease associated with contact-line crowding. The additional linear term empirically accounts for the renewed increase in line excess at higher concentrations. Its linear form is an approximation over the investigated concentration range rather than a separate Henry regime.

In the dilute limit, \Eq~\ref{eq:langmiur_line} reduces to the Henry regime $\Lambda=(L_1+L_2)c_0$, such that the line adsorption coefficient is $L=L_1+L_2$. In practice, $L_2\ll L_1$ and $L_1\simeq L$, with $L_2$ accounting for the high-concentration increase. The \emph{crossover concentration} $c^*$ provides an operational definition of the onset of contact-line crowding, while the exponent $\beta$ controls how sharply the isotherm departs from the  Henry regime.

The mRP isotherm accurately describes the nonmonotonic adsorption behavior of the alcohols and phenol over the investigated concentration ranges, as demonstrated by the solid lines in \Fig{fig:other_alc_line_ads}. The corresponding fit parameters are listed in Table~\ref{tab:parameters}.

A particularly insightful quantity is the crossover concentration, $c^*$, which approximately coincides with the local maximum of $\Lambda(c_0)$ and marks the onset of contact-line crowding. At this point, the line adsorption reaches $\Lambda(c^*)\approx1$--$2~\mathrm{nm}^{-1}$ ($\approx4~\mathrm{nm}^{-1}$ for C8). These values are consistent with the simple packing estimate of one amphiphile per molecular length $\Lambda(c^*)\simeq1/\sigma$, where $\sigma\approx0.5$ nm is the characteristic molecular size.

The onset of contact-line crowding depends strongly on amphiphile hydrophobicity. As shown in \Fig{fig:KL_vs_nc}c, $c^*$ decreases approximately exponentially with the alkyl chain length, $n_\mathrm{C}$, indicating that longer-chain amphiphiles crowd the contact line at progressively lower concentrations.

This dependence follows directly from the low-concentration scaling. Since contact-line crowding begins when $\Lambda(c^*)\simeq1/\sigma$, while $\Lambda\simeq Lc_0$ in the Henry regime (\Eq~\ref{eq:henry_line}), one obtains $c^* \simeq (\sigma L)^{-1}$.
Combining this expression with the exponential scaling of $L$ (\Eq~\ref{eq:exp_lin_ads}) gives
\begin{equation}
    c^*=c^{*(0)}\rme^{-\alpha_\mathrm{cl}n_\mathrm{C}},
    \label{eq:c*}
\end{equation}
The prediction contains no adjustable parameters apart from the prefactor $c^{*(0)}$ and is in excellent agreement with the simulation data (\Fig{fig:KL_vs_nc}c).
As we discuss below, $c^*$ provides the natural concentration scale for understanding amphiphile partitioning and crowding at the contact line.

\section{Discussion of Physical Implications}
The results establish the three-phase contact line as a distinct thermodynamic adsorption environment characterized by its own adsorption coefficient and characteristic crowding concentration. We now discuss the broader implications of these findings, focusing on the sensitivity to trace amphiphilic impurities, molecular partitioning across different length scales, and the consequences for line tension and heterogeneous nucleation.

\subsection{Contact-line enrichment and impurity sensitivity}

One of the most striking consequences of \Fig{fig:KL_vs_nc}c is the sensitivity of contact-line adsorption to trace amphiphilic impurities. Previous studies of surfactant-laden droplets have primarily considered deliberately added surfactants at relatively high concentrations, focusing on wetting, spreading, and Marangoni transport~\cite{isele2015smoothing, theodorakis2019molecular, theodorakis2019spreading, straub2021flow, reed2025neutron}. By contrast, our results show that substantial contact-line enrichment can occur even at impurity concentrations encountered in nominally clean aqueous systems.

Amphiphilic impurities are ubiquitous in aqueous systems~\cite{roger2012hydrophobic, maali2017viscoelastic}. Even ultrapure water can contain total organic carbon levels exceeding 1~\si{\ug}/L~\cite{uematsu2019impurity, zhang2021critical}, corresponding to organic molecule concentrations of approximately $10^{-7}$--$10^{-8}$~M, depending on molecular size. Remarkably, the exponential scaling in \Fig{fig:KL_vs_nc}c predicts $c^*\sim10^{-7}$~M for C12 and $\sim10^{-10}$~M for C16, comparable to or well below these impurity levels. Thus, even nominally clean droplets may possess substantially populated contact lines, with consequences for wetting behavior and line tension long before measurable changes occur at the neighboring interfaces.

Contact-line enrichment may also affect chemical processes. Three-phase contact lines are increasingly recognized as chemically active regions involved in electrochemical reactions, corrosion, crystallization, photocatalysis, and mineral dissolution~\cite{chen2021three, jimenez2021effect, weng2022engineering, wang2022identification, Wu2022}. The accumulation of amphiphilic reactants, intermediates, or impurities may alter local reaction kinetics and nucleation pathways.

As an example, consider a bimolecular reaction between reactants $A$ and $B$, with a local reaction rate proportional to the product of their concentrations. For a spherical-cap droplet with base radius $a$, the reaction rate at the water--solid interface scales as $R_\mathrm{ws}\propto \pi a^2 \Gamma_{\mathrm{ws},A}\Gamma_{\mathrm{ws},B}$.
At the contact line, we assume a reaction zone of effective width $\sigma\approx 0.5$~nm, where the effective surface concentrations are $\Lambda_A/\sigma$ and $\Lambda_B/\sigma$. Assuming comparable intrinsic reaction kinetics, the contact-line contribution scales as $R_\mathrm{cl}\propto 2\pi a\sigma(\Lambda_A/\sigma)(\Lambda_B/\sigma)$.
In the Henry regime, the ratio of the two contributions is therefore
$R_\mathrm{cl}/R_\mathrm{ws}\sim 2L_A L_B/(\sigma K_{\mathrm{ws},A}K_{\mathrm{ws},B}a)$.

For reactants with similar adsorption properties, the contributions from the contact line and surface become comparable at a characteristic droplet radius $a^*\sim 2(L/K_\mathrm{ws})^2/\sigma\propto\exp[2(\alpha_\mathrm{cl}-\alpha_\mathrm{ws})n_\mathrm{C}]$. This radius increases approximately 1.8-fold per additional carbon atom, reaching about 60 nm for C5, 350 nm for C8, and an extrapolated 3.5 $\mu$m for C12.

Although this estimate assumes that adsorption excesses represent reactive molecular populations, it suggests that contact-line enrichment could substantially influence chemical reactions even in micrometer-sized droplets.

\subsection{Competition between adsorption reservoirs}

Adsorption partitions amphiphilic molecules among three thermodynamically distinct reservoirs: the bulk liquid, interfaces, and three-phase contact lines. The relative population of these reservoirs depends not only on the adsorption affinity of the amphiphiles, but also on the geometry and characteristic size of the system.
The partitioning follows directly from the Gibbs decomposition (eq~\ref{eq:line_ads}). Although introduced above for the cylindrical simulation geometry, the same framework applies to any geometry. 

To illustrate this partitioning for a specific geometry, we now consider a sessile spherical-cap water droplet. For a droplet with radius of curvature $r$, contact angle $\theta$, and base radius $a=r\sin\theta$, the water--vapor and water--solid interfacial areas are $A_\mathrm{wv}=2\pi r^2(1-\cos\theta)$ and $A_\mathrm{ws}=\pi a^2$, while the droplet volume is $V=(\pi/3)r^3(2+\cos\theta)(1-\cos\theta)^2$. The corresponding fractions of amphiphiles residing at the contact line, interfaces, and in the bulk are
$f_\mathrm{cl}={2\pi a\Lambda}/{N}$, $ f_{A}={(\Gamma_\mathrm{wv}A_\mathrm{wv}+\Gamma_\mathrm{ws}A_\mathrm{ws})}/{N}$, and $f_{V}={V c_0}/{N}$, respectively. For simplicity, we restrict the analysis to the dilute adsorption regime.

\begin{figure}[h]
  \includegraphics[width=0.495\linewidth]{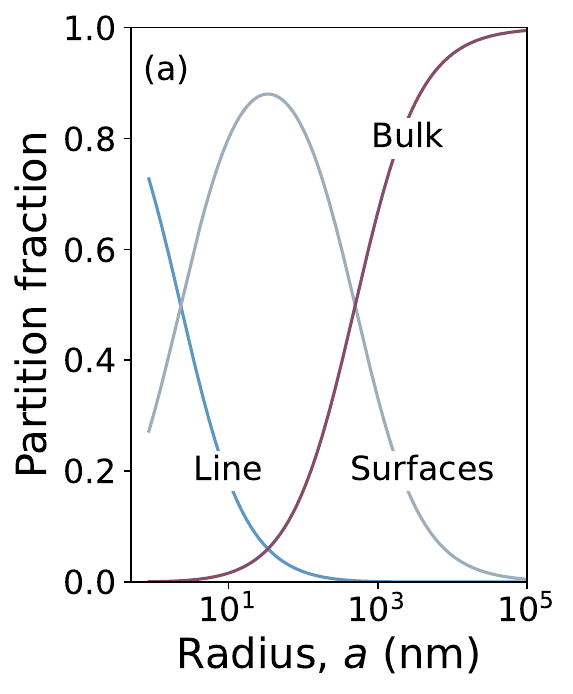}
  \includegraphics[width=0.495\linewidth]{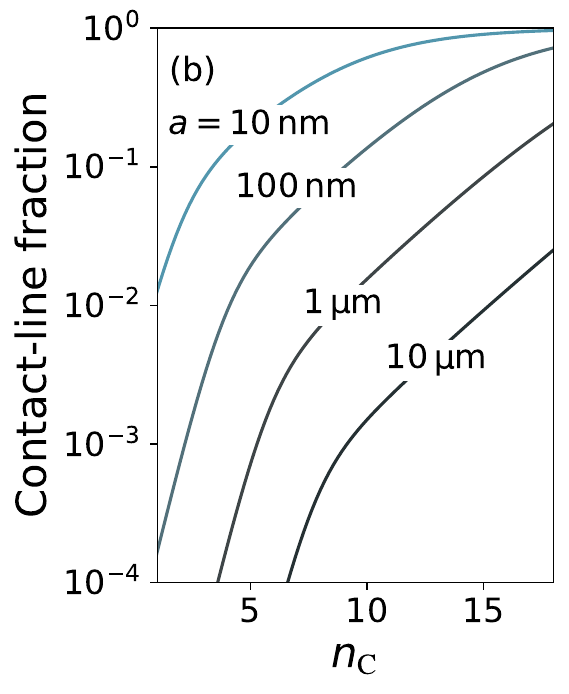}
  \caption{\textbf{Competition between adsorption reservoirs.}
  (a) Predicted partition fractions for pentanol (C5) in a sessile spherical-cap droplet on a hydrophobic substrate ($\theta=120^\circ$) as a function of droplet base radius. The partition fractions are calculated using the adsorption coefficients obtained from the MD simulations. (b) Predicted partition fraction of amphiphiles residing at the contact line as a function of alkyl chain length for droplets of different base radii $a$.}
  \label{fig:partition}
\end{figure}

The partitioning changes dramatically with droplet size. As shown in \Fig{fig:partition}a for pentanol (C5), the contact line is the dominant adsorption reservoir for droplets with base radii up to approximately 5~nm. As the droplet becomes larger, adsorption is progressively transferred to the interfaces, while bulk partitioning dominates only for droplets approaching the micrometer scale. These three regimes arise directly from geometry: the contact line scales with the droplet circumference ($\sim a$), the interfaces with area ($\sim a^2$), and the bulk with volume ($\sim a^3$).

The crossover in prevalence between contact-line and surface partitioning occurs at a droplet size $a\sim L/K_\mathrm{ws}$ and therefore depends strongly on amphiphile type. Because $L$ increases more rapidly with chain length than $K_\mathrm{ws}$, the crossover shifts to progressively larger droplets for longer alkyl chains, scaling as
$a \sim \exp[(\alpha_\mathrm{cl}-\alpha_\mathrm{ws})n_\mathrm{C}]$.

To illustrate this trend, we combine the exponential scaling relations for $L$, $K_\mathrm{wv}$, and $K_\mathrm{ws}$  to calculate the contact-line partitioning fraction as a function of alkyl chain length (\Fig{fig:partition}b).
 The fraction of amphiphiles residing at the contact line increases rapidly with chain length, such that contact-line partition remains significant even for micron-sized droplets when sufficiently long amphiphiles are present. For example, our model predicts that several percent of C16 molecules in a micron-sized droplet reside at the contact line.

These estimates demonstrate that the contact line cannot, in general, be regarded as a negligible geometric feature. Despite occupying only a vanishingly small fraction of the droplet volume, it can become the dominant adsorption reservoir over a surprisingly broad range of droplet sizes and amphiphile chemistries. This effect is expected to be particularly important in systems containing large densities of contact lines, such as porous, fibrous, and wet granular materials, Pickering emulsions, rough or textured surfaces, and collections of microscopic droplets such as those encountered during condensation or spraying.

\subsection{Adsorption-induced reduction of line tension}
A direct consequence of contact-line adsorption is a reduction of the line tension. Since the line tension, $\tau$, is the excess free energy per unit length of the three-phase contact line, adsorption of amphiphiles lowers its value, analogous to the reduction of surface tension by surfactants.

The connection follows from the Gibbs adsorption relation for the contact line,
 $\mbox{d} \tau= -\Lambda\,\mbox{d}\mu $,
which links changes in line tension directly to the line adsorption~\cite{rowlinson2013molecular}.  A more general expression contains an additional geometric term, $c_\theta\mbox{d}\theta$, associated with changes in the equilibrium contact angle that is required for thermodynamic consistency~\cite{djikaev2004geometric}.  Our previous simulations for the present substrate indicate that this contribution is negligibly small~\cite{Kanduc2017b}, and it is therefore neglected here.
At sufficiently low concentrations, where amphiphile--amphiphile interactions in the bulk are negligible, the solution behaves ideally, with chemical potential $\mu = \mu_{\mathrm{ex}} + k_{\mathrm{B}}T \ln c_0$ and an approximately constant excess term $\mu_{\mathrm{ex}}$~\cite{Staniscia22a,kanduc2023interface}. Integration then yields the change in line tension associated with adsorption,
\begin{equation}\label{eq:dtau}
    \Delta \tau = - k_\mathrm{B} T\, \int_0^{c_0} \Lambda(c_0') \frac{\mbox{d}c_0^{\prime}}{c_0'}
\end{equation} 

Although the Gibbs relation has been established theoretically for decades~\cite{widom2004adsorption, djikaev2004geometric, widom2006models, rowlinson2013molecular}, quantitative predictions of line tension based directly on molecular adsorption isotherms have remained largely unexplored. To our knowledge, only one previous molecular simulation study reported preferential accumulation of a solute (CO$_2$) at a three-phase contact line and related it to changes in line tension~\cite{bey2021carbon}. Here, we show that line adsorption follows systematic trends across chemically diverse amphiphiles, providing a molecular thermodynamic foundation for line-active behavior at three-phase contact lines.

\begin{figure}
  \includegraphics[width=0.8\linewidth]{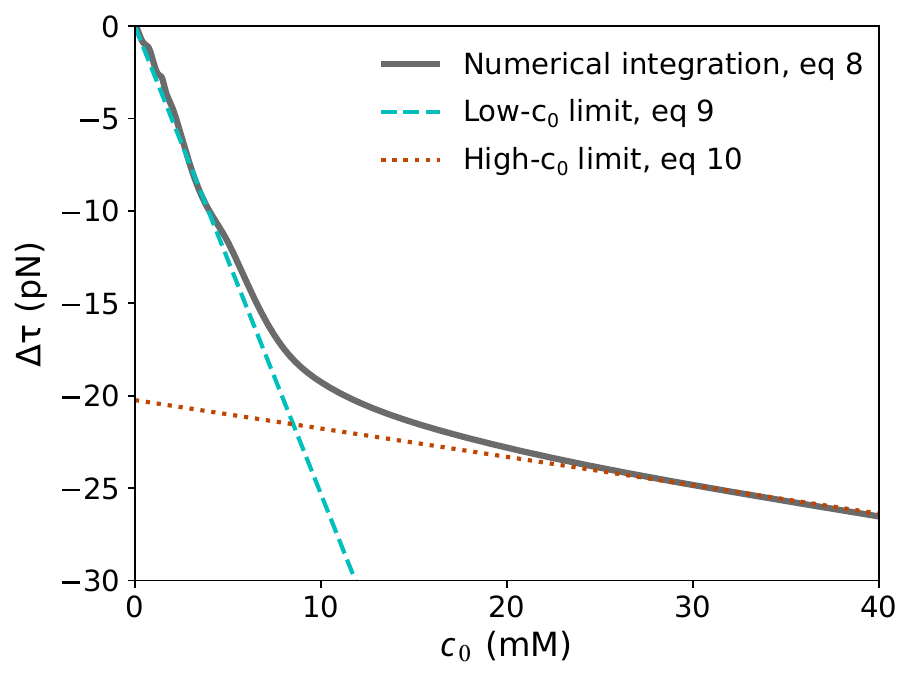}
  \caption{\textbf{Adsorption-induced reduction of the line tension.} Reduction of the line tension, $\Delta\tau$, due to amphiphile adsorption at the contact line as a function of bulk concentration for pentanol (C5). The solid line shows the numerical evaluation of the Gibbs adsorption relation (eq~\ref{eq:dtau}), while the dashed and dotted lines indicate the asymptotic low- and high-concentration limits (\Eqs~\ref{eq:dtau_low} and \ref{eq:dtau_high}), respectively.}
  \label{fig:ratio_deltatau}
\end{figure}

Using a spline interpolation of the simulated $\Lambda(c_0)$, we integrate eq~\ref{eq:dtau} numerically. The resulting line-tension reduction for C5 is shown in \Fig{fig:ratio_deltatau}. The line tension initially decreases linearly with concentration and then crosses over to a weaker concentration dependence once the contact line becomes crowded above $\sim c^*$. Similar behavior is obtained for other amphiphiles studied (NB is excluded because of its weaker amphiphilic character, see Section S5 in the SI).

Simple analytical expressions can be obtained in the dilute and crowded limits. For the low-concentration limit ($c_0\ll c^*$), where $\Lambda\simeq Lc_0$, one finds
\begin{equation}
    \label{eq:dtau_low}
    \Delta \tau = - k_\mathrm{B} T L c_0
\end{equation}
whereas for the high-concentration limit ($c_0\gg c^*$), one obtains
\begin{equation}
\label{eq:dtau_high}
\Delta\tau=\Delta\tau^*-k_{\mathrm{B}}TL_2c_0,
\end{equation}
with
\begin{equation}
\label{eq:dtau*}
\Delta\tau^*=
-\frac{\pi k_{\mathrm{B}}T}{\beta\sin(\pi/\beta)}Lc^*.
\end{equation}
Here, $\Delta\tau^*$ comes from the integration of the first term of the mRP relation up to infinite concentration, using the approximation $L_1\simeq L$.
Both limiting predictions are also shown in \Fig{fig:ratio_deltatau}.

The crossover line tension, $\Delta\tau^*$, is of the same order as the reduction at the onset of contact-line crowding, $\Delta\tau^*\sim\Delta\tau(c^*)$. All amphiphiles studied yield comparable values, $\Delta\tau^*\sim-20$~pN (Table~\ref{tab:parameters}). This apparent universality follows naturally from molecular packing. Since contact-line crowding begins when $Lc^*\sim1/\sigma$, where $\sigma\approx0.5$~nm is the characteristic molecular size, eq~\ref{eq:dtau*} predicts
$\Delta \tau^{*} \simeq -k_{\mathrm{B}}T/\sigma\sim-10$ pN, up to a numerical prefactor. 

Remarkably, this reduction is comparable to the intrinsic magnitude of the line tension itself, $\tau\sim\Delta\tau^*$. While experimental values reported in the literature span several orders of magnitude and often even differ in sign~\cite{drelich1992line, duncan1995correlation, pompe2000three, guillemot2012activated, leelamanie2012drop, law2017line, matsubara2018finite, zhao2019resolving}, measurements on atomistically smooth and chemically clean systems, together with molecular simulations, generally place line tensions in the piconewton range~\cite{drelich1996significance, weijs2011origin, guillemot2012activated, werder2003water,sergi2012molecular, khalkhali2017wetting, Kanduc2017b}.
This magnitude is naturally explained by the scaling relation $\tau\sim\gamma\sigma$, where $\gamma$ is the surface tension and $\sigma$ is a molecular length scale~\cite{weijs2011origin, guillemot2012activated}.

This comparison has an important implication. Because the crossover concentration $c^*$ decreases exponentially with amphiphile size, the concentrations required to modify the line tension by its intrinsic magnitude fall below the micromolar or nanomolar range for long-chain amphiphiles (\Fig{fig:KL_vs_nc}c). These concentrations overlap with the practical purity limit of aqueous systems. Consequently, trace amphiphilic impurities may strongly affect line tension and interfacial phenomena even in nominally clean experiments. In particular, adsorption at the contact line may modify contact-angle hysteresis by changing the free-energy barriers for contact-line motion~\cite{giacomello2016wetting}. 
Line tension also affects heterogeneous bubble nucleation, where it contributes a free-energy term of $2\pi a\tau$, with $a$ being the bubble base radius.

For a critical base radius of $a\approx1$ nm, a line-tension reduction of $\Delta\tau=-20$ pN lowers the nucleation barrier by approximately $30\,\kB T$.
This reduction in the nucleation barrier also lowers the critical gas supersaturation, $S_\mathrm{c}$, required for appreciable bubble nucleation. As shown in Section S6 in the SI, a perturbative analysis predicts $\Delta S_\mathrm{c}/S_\mathrm{c}\propto S_\mathrm{c}\Delta\tau$, indicating that the effect of line tension is particularly pronounced at high supersaturations, such as those encountered during electrochemical gas generation. For example, experiments on H$_2$ nanobubble nucleation at Pt nanoelectrodes reported $S_\mathrm{c}\approx300$~\cite{chen2014electrochemical, german2016electrochemistry}. Under these conditions, a line-tension reduction of $\Delta\tau=-20$ pN is predicted to lower $S_\mathrm{c}$ by approximately 25\%. These studies also found that surfactants facilitate bubble nucleation, an effect attributed mainly to reduced surface tension. Our results suggest that contact-line adsorption could provide an additional mechanism.

Line tension may likewise influence heterogeneous cavitation. For a smooth hydrophobic surface, the same reduction of $\Delta\tau=-20$ pN is estimated to decrease the magnitude of the cavitation pressure by nearly 50\%, from $-60$ to approximately $-30$ MPa. However, because cavitation is a rapid process, amphiphiles may not have sufficient time to adsorb at the emerging contact line, potentially limiting this effect in practice.

\section{Conclusions}

The present work establishes the three-phase contact line on a substrate as a distinct thermodynamic adsorption environment, rather than merely the geometric intersection of two interfaces. Amphiphilic molecules preferentially accumulate at this lower-dimensional region, where their adsorption can be stronger than at either neighboring interface, thereby acting as line-active agents (lineactants). We quantify adsorption through the line adsorption coefficient and the characteristic concentration for contact-line crowding.

Despite their different molecular structures, all six amphiphiles follow a common scaling relation between line and water--solid adsorption in the Henry regime, suggesting that contact-line affinity is largely governed by affinity for the solid surface. At higher concentrations, the adsorption isotherms become molecule-specific, although the surfactant-like molecules (alcohols and phenol) exhibit similar nonmonotonic behavior.

These findings have important physical consequences. Because the onset of contact-line crowding shifts rapidly to lower concentrations with increasing amphiphile hydrophobicity, even trace impurities may substantially populate contact lines under nominally clean conditions. Such enrichment can alter molecular partitioning within droplets and modify line tension by an amount comparable to its intrinsic magnitude, with potential consequences for wetting, heterogeneous nucleation, and interfacial reactions. While our analysis focuses on atomistically smooth hydrophobic substrates, understanding how chemical heterogeneity, surface roughness, and hydrophilicity influence contact-line adsorption remains an important direction for future work.
\\

\section{Methods}
\label{sec:methods}

\subsection*{Simulations}

We performed MD simulations of amphiphilic molecules adsorbing on a model substrate in water. Water was described by the SPC/E model~\cite{berendsen1987}, while the amphiphiles and substrates were modeled with the GROMOS force field~\cite{Oostenbrink2004}. The amphiphile parameters were obtained from the ATB repository~\cite{Malde2011}.

The substrate was atomistically smooth and based on our previous model~\cite{Kanduc2014, Kanduc2017b, Staniscia22a}. 
It consisted of a bilayer of hexagonally packed aliphatic chains (10 carbon atoms long) terminated by hydroxyl ($-\mathrm{OH}$) head groups with a surface density of $4.3$~nm$^{-2}$. Its hydrophobicity was tuned by rescaling the atomic partial charges, thereby varying the polarity of the terminal hydroxyl groups. In the present study, we used the parameterization corresponding to a water contact angle of $\approx 120^\circ$.

The simulation box had a height of 10--12~nm ($z$-direction) and a depth of 5.2~nm ($y$-direction). The dimensions in the $x$-direction were 9.0 nm (in most cases), 13.5 nm (at high amphiphile concentrations to accommodate the stronger droplet spreading), and 18.0~nm (for the highest C5 concentration).   

The simulations were performed with GROMACS 2023 using a time step of 2~fs in the NVT ensemble with periodic boundary conditions in all three dimensions. The temperature was maintained at 300~K with the velocity-rescaling thermostat~\cite{Bussi07} using a coupling time constant of 0.1~ps. Electrostatic interactions were treated with the particle--mesh Ewald method~\cite{PME1, PME2} using a real-space cutoff of 0.9~nm, while Lennard-Jones interactions were truncated at the same distance. These simulation settings were chosen to remain consistent with our previous studies, in which the substrate model was extensively validated~\cite{Kanduc2014, Kanduc2017b, Staniscia22a}.

For each system, we performed three independent simulations. Most production runs lasted 500~ns, except for methanol and propanol at the highest concentrations, for which 100~ns was sufficient to sample the equilibrium behavior, and octanol, for which 1~$\mu$s simulations were required because of its slower dynamics. The first 20\% of each trajectory was discarded as equilibration.

The MD input files used in this study are available in a Zenodo repository (see the Data Availability section).

\subsection*{Droplet partitioning}

\begin{figure}[h]
\includegraphics[width=0.65\linewidth]{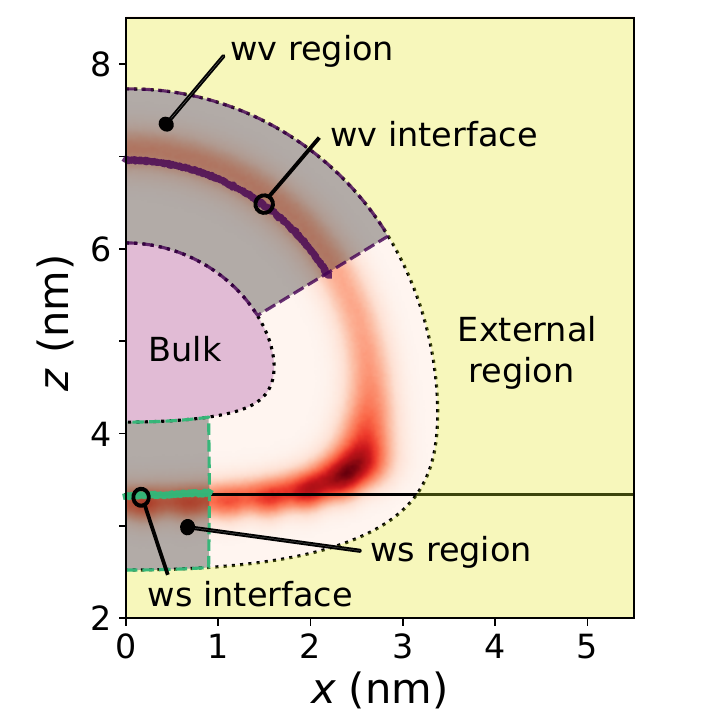}
  \caption{\textbf{Definition of the regions used for adsorption analysis.} Partitioning of the representative droplet shown in \Fig{fig:densities}c into the bulk, interfacial, and external regions used to evaluate adsorption. The bulk region is shaded purple, the water--vapor and water--solid interfacial regions are shaded gray, and the external region is shaded yellow.}
  \label{fig:comparts}
\end{figure}

To quantify adsorption at the water--vapor interface, the water--solid interface, and the three-phase contact line, the simulation domain was partitioned into the regions illustrated in \Fig{fig:comparts}. The central part of the droplet defines the bulk region, from which the bulk amphiphile concentration,  $c_0$, was determined. The surrounding water--vapor and water--solid interfacial regions were then identified to evaluate the corresponding surface adsorptions, $\Gamma_{\mathrm{wv}}$ and $\Gamma_{\mathrm{ws}}$. These quantities, together with the interfacial areas of the cylindrical reference geometry, were substituted into \Eq~\ref{eq:line_ads} to obtain the line adsorption, $\Lambda$.

The complete procedure is described in Section S7 in the SI, and the corresponding Python implementation is provided in the Data Availability section.

%\subsection*{Fits}
%To obtain the most precise values of $L$ we extracted them from a fit, at low $c_0$, of the relation~\ref{eq:langmiur_line}. We then fitted the parameters of the mRP isotherm~\ref{eq:langmiur_line} to the whole set of data points, with the constraint that $L=L_1+L_2$, which left only three independent parameters. We used this two-step procedure since data follow very well the linear laws at small concentration, while the agreement with the mRP isotherm are less precise, and this procedure avoids that these deviations affect and spoil the fit of the value of $L$.

\section*{CRediT authorship contribution statement}
{\bf Fabio Staniscia}: Conceptualization, Methodology, Formal analysis,  Investigation, Writing – original draft, Visualization. {\bf Matej Kandu\v{c}}: Conceptualization, Investigation, Writing – review \& editing, Funding acquisition.

\section*{Declaration of competing interest}
The authors declare no competing financial interest.

\section*{Data availability}
The MD simulation input files used in this study are publicly available in the Zenodo repository at \url{https://doi.org/10.5281/zenodo.21683705}.
The Python scripts used for data analysis are available at
\url{https://github.com/hckq/analysis_md/tree/main}.

\section*{Acknowledgments} 
Authors acknowledge financial support from the Slovenian Research and Innovation Agency ARIS (contracts P1-0055, J1-4382, J1-70038).

\setlength{\bibsep}{1pt}
\small
\bibliography{bibliography}

\end{document}

\singlespacing
\pagenumbering{arabic}
\renewcommand{\thepage}{S\arabic{page}}  
\noindent

\parindent=0cm
\setlength\arraycolsep{2pt}

\maketitle

%\maketitle
\setlength\arraycolsep{2pt}

\section{Adsorbed amphiphile orientation}

\label{sec:orientation}
\subsection{Orientation at the contact line}

\begin{figure}[h]
\centering
\includegraphics[width=0.6\linewidth]{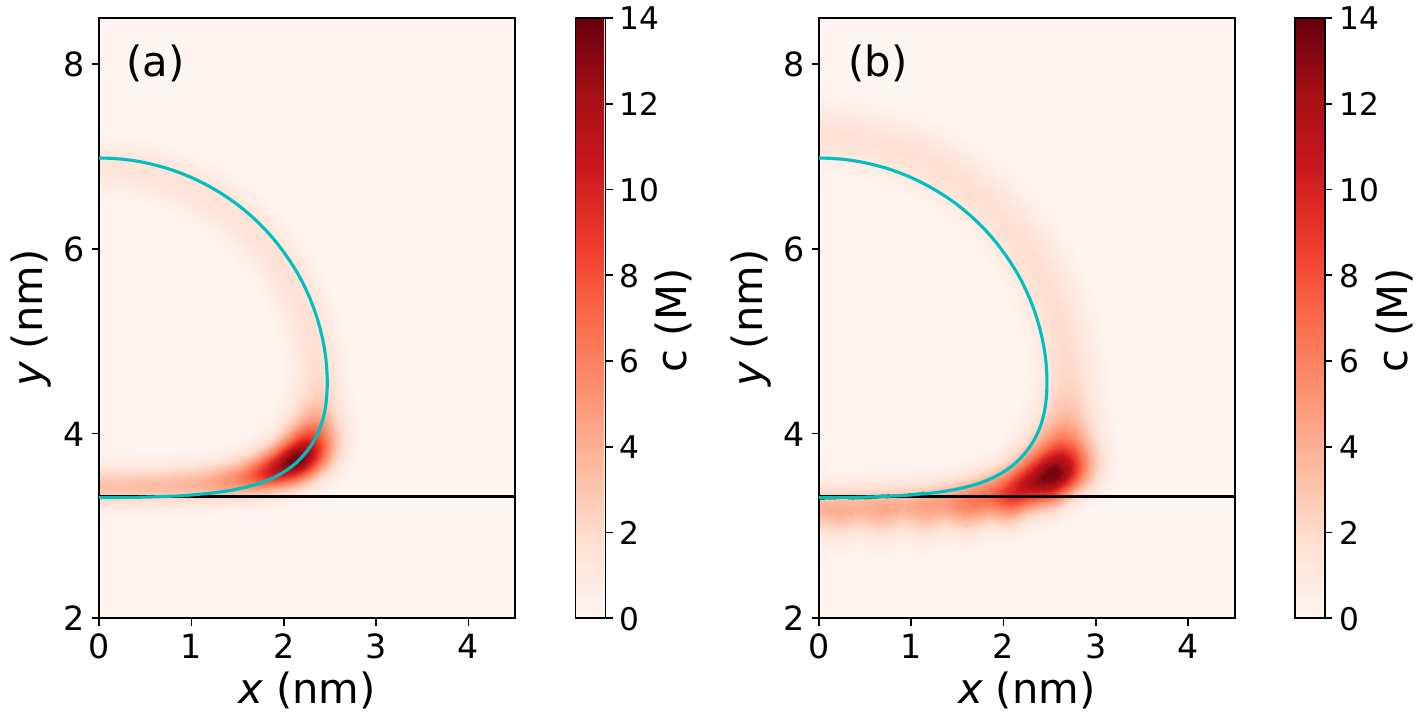}
\caption{Density maps of pentanol (C5) at a bulk concentration of $c_0\approx2.0$~mM, calculated using the positions of (a) the hydroxyl oxygen atom and (b) the terminal carbon atom of the hydrocarbon chain, rather than the molecular center of mass.}
\label{fig:carb_oxy_comparison}
\end{figure}

We first examine the spatial arrangement of amphiphiles adsorbed at the contact line. \Fig~\ref{fig:carb_oxy_comparison} shows the concentration maps of the hydroxyl oxygen and terminal carbon atoms of pentanol (C5). The hydroxyl groups are predominantly located near the droplet interface, whereas the hydrocarbon tails extend outward, largely outside the droplet.

\begin{figure}[h]
\includegraphics[width=0.6\linewidth]{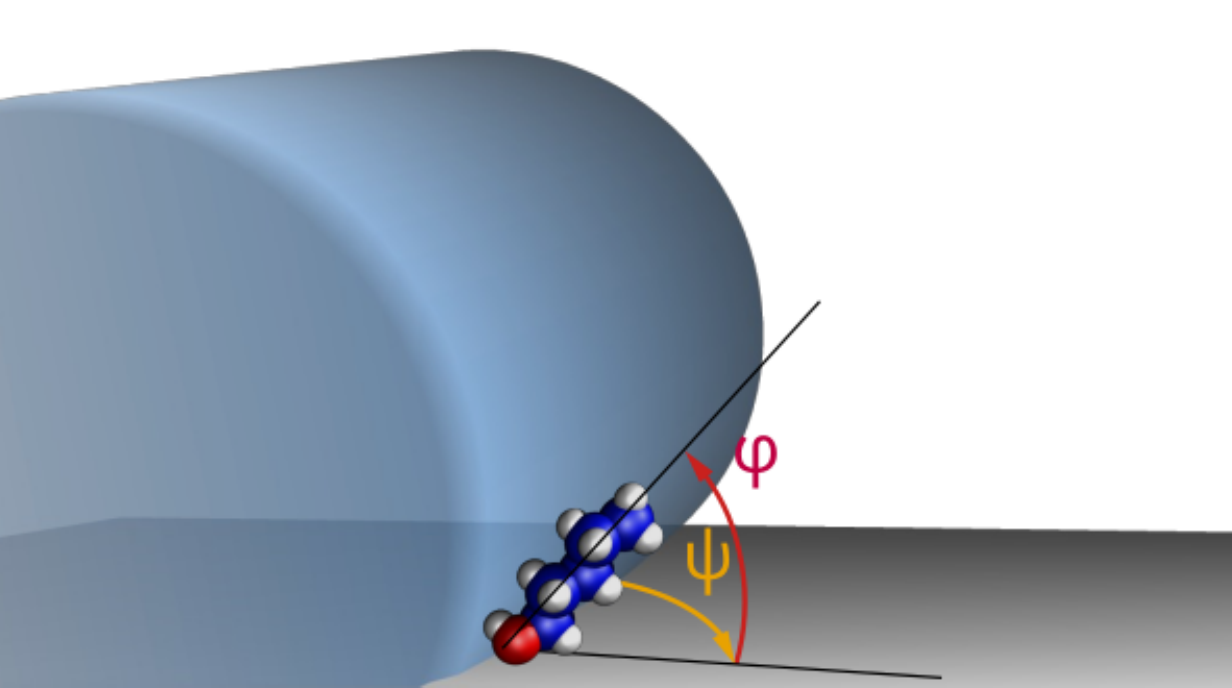}
\caption{Definition of the tilt angle $\varphi$ and azimuthal angle $\psi$ describing the orientation of an amphiphilic molecule adsorbed at the contact line.}
\label{fig:schem_angles}
\end{figure}

To quantify the molecular orientations, we define a vector pointing from the hydrophilic headgroup (O or N atom) to the carbon atom farthest from the headgroup. As illustrated schematically in \Fig~\ref{fig:schem_angles}, the orientation of this vector is described by two angles. The tilt angle, $\varphi$, measures the inclination of the molecular axis with respect to the substrate plane, whereas the azimuthal angle, $\psi$, specifies its orientation relative to the contact-line direction. Thus, $\psi=0$ or $\psi=\pi$ corresponds to a molecule aligned along the contact line, whereas $\psi=\pi/2$ corresponds to a perpendicular alignment.

\begin{figure}[h]
\includegraphics[width=0.9\linewidth]{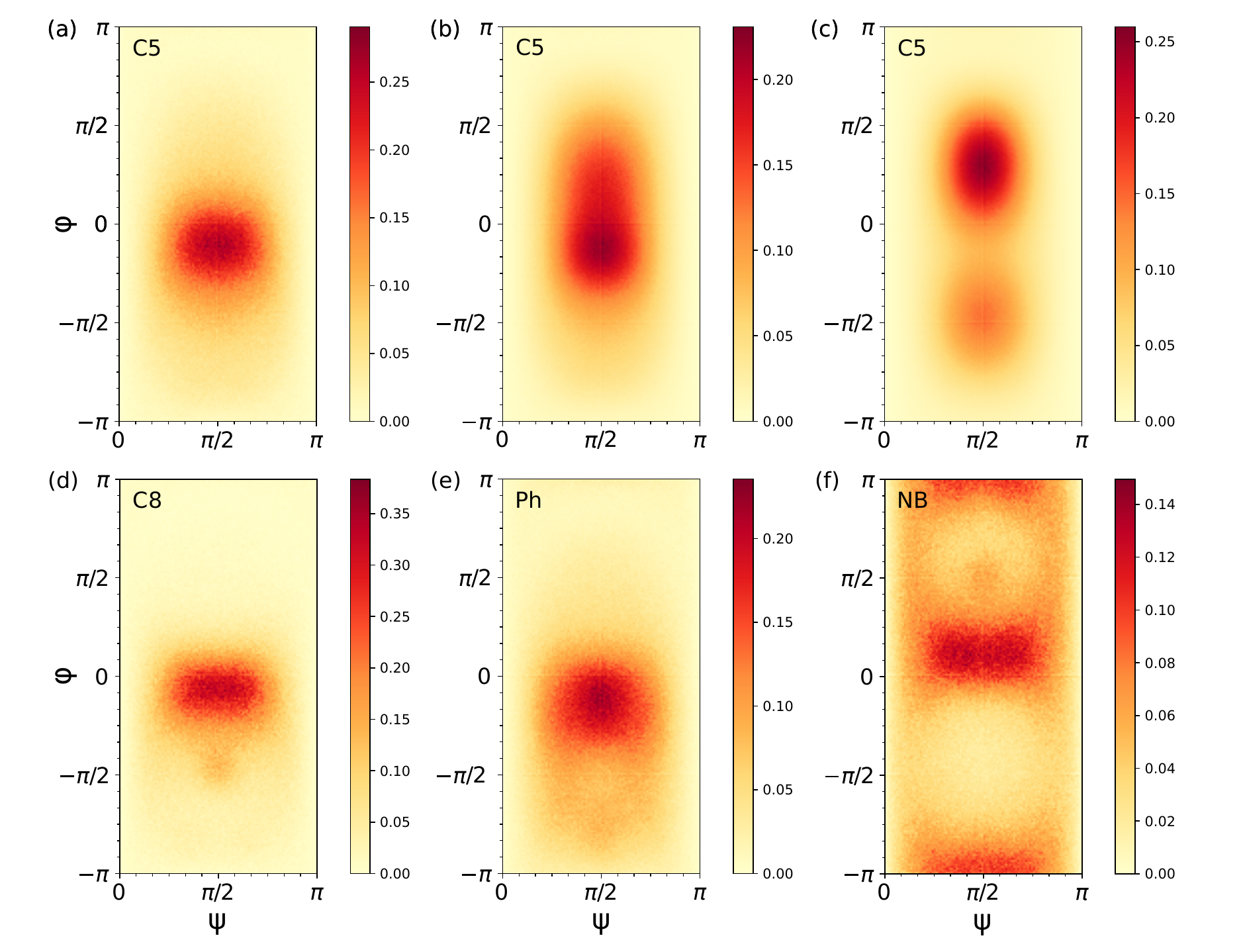}
\caption{Orientation distributions of amphiphiles adsorbed at the contact line for pentanol at increasing concentrations: (a)~$c_0 \approx 0.85$~mM, (b) $8.9$~mM, and (c) $68$~mM,  (d) octanol at $c_0\approx0.0040$~mM, (e) phenol at $c_0 \approx 2.8$~mM, and (f) nitrobenzene at  $c_0 \approx0.35$~mM.
The color scales represent normalized probability densities.}
\label{fig:orientation}
\end{figure}

The orientation distributions for C5, C8, phenol (Ph), and nitrobenzene (NB) are shown in \Fig~\ref{fig:orientation}. The distributions of C5, C8, and Ph exhibit a pronounced maximum near $\psi=\pi/2$ and $\varphi\approx0$, indicating that the molecules preferentially lie nearly flat on the substrate, with their hydrocarbon tails pointing radially outward, perpendicular to the contact line. 

Nitrobenzene exhibits qualitatively different behavior. The distribution of $\varphi$ is bimodal, with maxima near $0$ and $\pi$, while $\psi$ remains broadly distributed. This indicates that nitrobenzene lies predominantly flat on the substrate but has little preference for the orientation of the NO$_2$ group relative to the contact line. Overall, nitrobenzene displays substantially weaker orientational ordering than the linear alcohols and phenol, consistent with its comparatively weak amphiphilic character.

\subsection{Orientation at the water--solid interface}

Since amphiphiles at the contact line preferentially lie nearly flat on the substrate, we next examine whether the same orientational preference persists at the water--solid interface away from the contact line.

\begin{figure}
\includegraphics[width=0.5\linewidth]{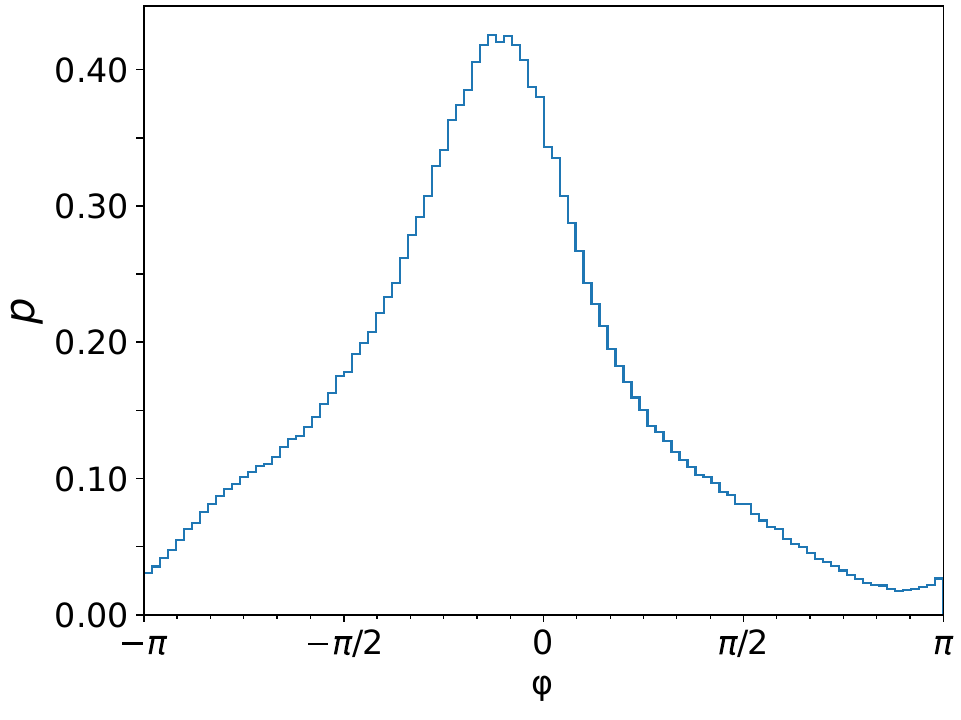}
\caption{Distribution of the tilt angle $\varphi$ for pentanol molecules adsorbed at the water--solid interface. }
\label{fig:distr_ws}
\end{figure}

Away from the contact line, the azimuthal angle $\psi$ is no longer biased by the contact-line direction and is therefore uniformly distributed. Consequently, only the tilt-angle distribution $\varphi$ is informative. As shown in \Fig~\ref{fig:distr_ws}, the distribution exhibits a pronounced maximum at $\varphi\approx0$, indicating that the amphiphiles also adopt nearly flat configurations on the substrate.  Thus, the preferred molecular tilt at the contact line primarily reflects the interaction with the hydrophobic substrate, whereas the water--vapor interface mainly determines the azimuthal orientation by breaking the in-plane rotational symmetry.

The close similarity between the contact-line and water--solid orientations therefore suggests that the substrate dominates the orientational ordering at the contact line. This conclusion is consistent with the observation in the main text that line adsorption is influenced more strongly by the water--solid interface than by the water--vapor interface.

\section{Surface adsorption}
\label{sec:surf_ads}

\begin{figure*}
  \includegraphics[width=0.33\linewidth]{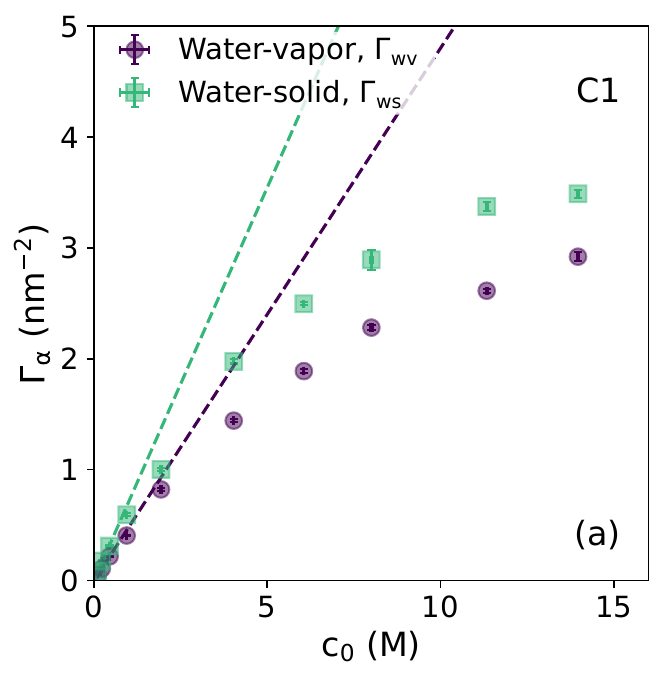}
  \includegraphics[width=0.312\linewidth]{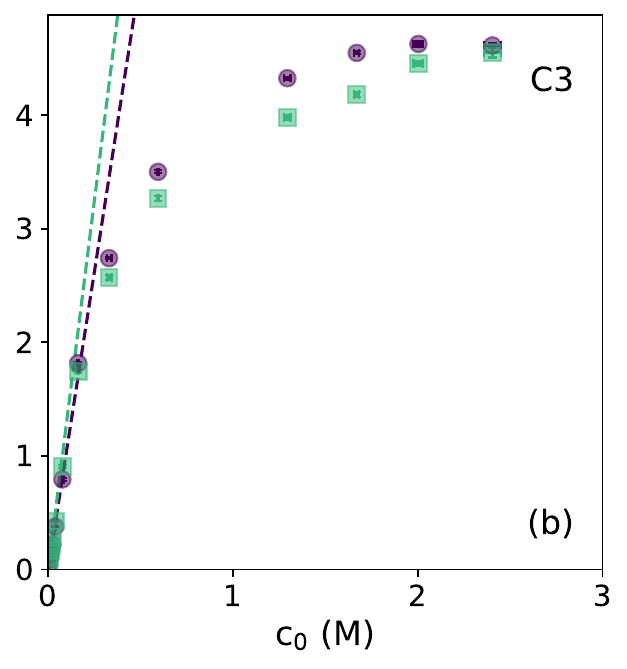}
      \includegraphics[width=0.32\linewidth]{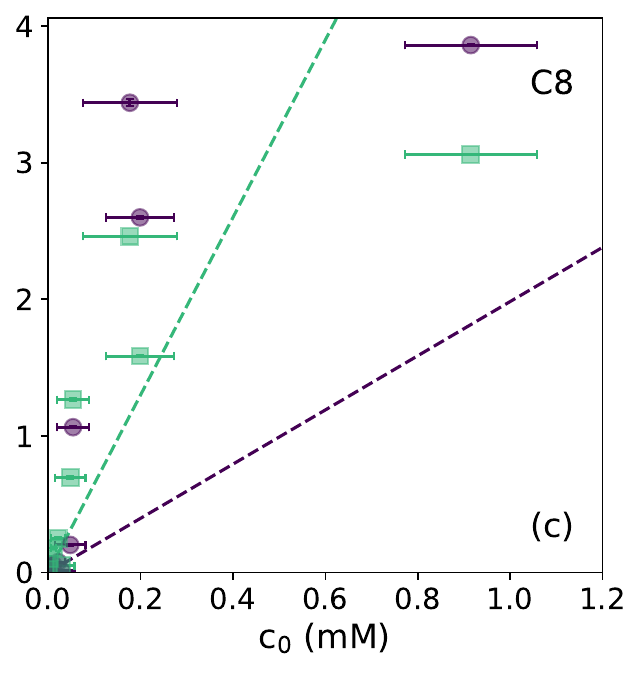}   \\        \includegraphics[width=0.33\linewidth]{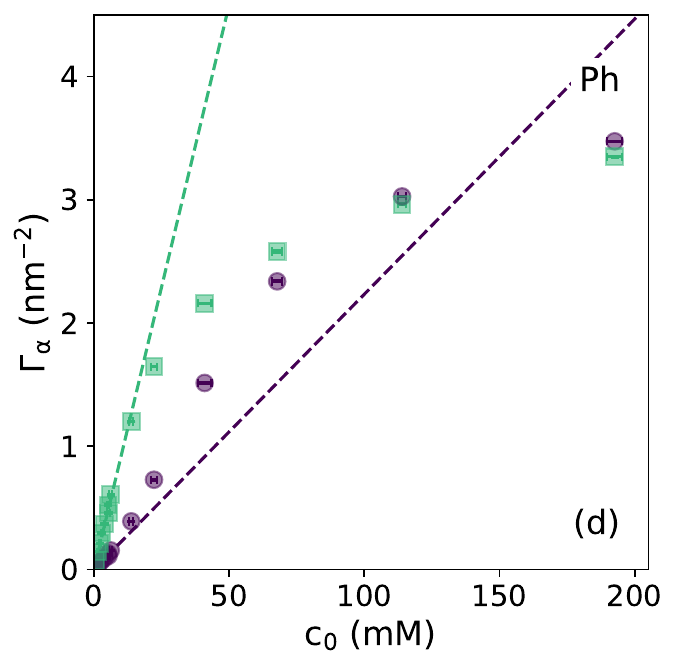} 
      \includegraphics[width=0.315\linewidth]{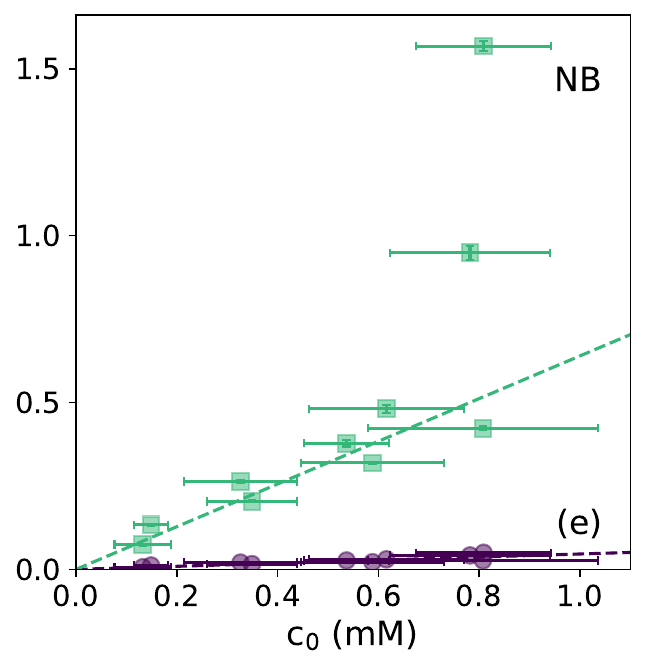}
  \caption{Surface adsorption isotherms for all amphiphiles investigated in this work except pentanol (C5), which is shown in the main text. Dashed lines represent Henry's law fits in the low-concentration regime.}
  \label{fig:alc_surf_ads}
\end{figure*}

Figure~\ref{fig:alc_surf_ads} shows the surface adsorption isotherms for the water--vapor and water--solid interfaces that are not included in the main text. The behavior is consistent with previous simulations of flat water slabs~\cite{Staniscia22a} and spherical and cylindrical nanodroplets~\cite{Staniscia25}, exhibiting an initial Henry regime followed by saturation at higher concentrations.

The low-concentration regime is fitted with Henry's law,
\begin{equation}
    \Gamma_\mathrm{wv} = K_\mathrm{wv} \, c_0
    \quad\textrm{and}\quad 
    \Gamma_\mathrm{ws} = K_\mathrm{ws} \, c_0
\end{equation}
where $\mathrm{wv}$ and $\mathrm{ws}$ stand for water--vapor and water--solid, respectively. The resulting adsorption coefficients, $K_\mathrm{wv}$ and $K_\mathrm{ws}$, are reported in Table~1 of the main text.

\begin{figure}
\centering
     \includegraphics[width=0.4\linewidth]{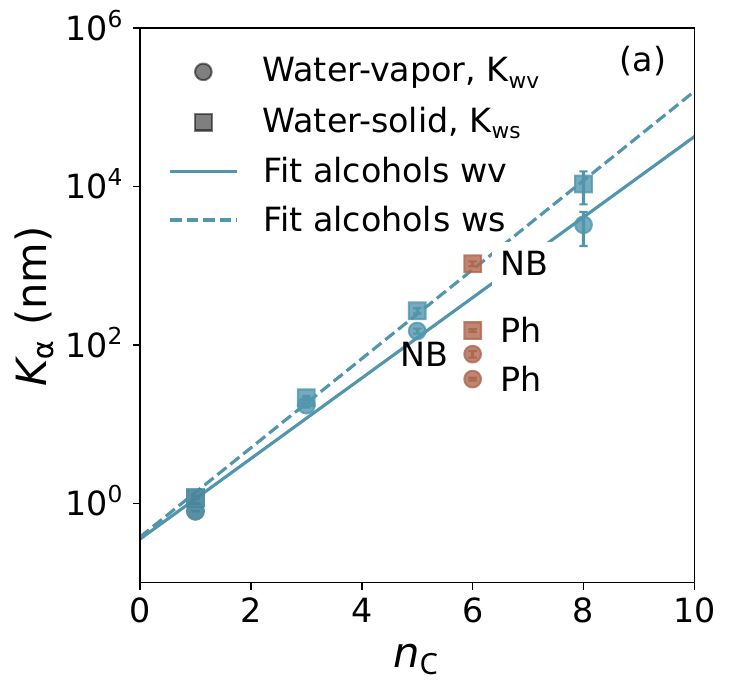}
  \includegraphics[width=0.39\linewidth]{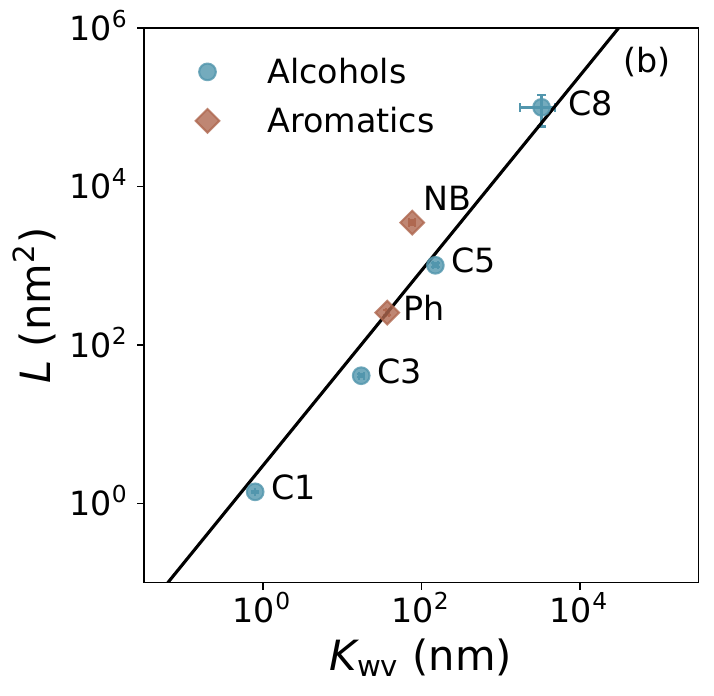}
  \caption{(a) Surface adsorption coefficients as a function of the number of carbon atoms. The lines show exponential fits of \Eq~\ref{eq:Knc}. (b) Line adsorption coefficient, $L$, plotted against the water--vapor surface adsorption coefficient, $K_{\mathrm{wv}}$. The solid line shows the prediction of \Eq~\ref{eq:scaling3}.}
  \label{fig:L_vs_K_wv}
\end{figure}

Figure~\ref{fig:L_vs_K_wv}a shows the dependence of the adsorption coefficients on the number of carbon atoms, $n_{\mathrm C}$. As in previous studies, the adsorption coefficients of linear alcohols follow an exponential dependence,
\begin{equation}
K_\mathrm{wv} \simeq K_\mathrm{wv}^{(0)}\rme^{\alpha_\mathrm{wv} n_{\mathrm C}}
\quad\textrm{and}\quad 
K_\mathrm{ws} \simeq K_\mathrm{ws}^{(0)}\rme^{\alpha_\mathrm{ws} n_{\mathrm C}}
\label{eq:Knc}
\end{equation}

with  $\alpha_{\mathrm{wv}}\approx1.17$ and $\alpha_{\mathrm{ws}}\approx1.29$.
To compare line adsorption with adsorption at the water--vapor interface, \Fig~\ref{fig:L_vs_K_wv}b plots the line adsorption coefficient, $L$, as a function of the water--vapor adsorption coefficient, $K_{\mathrm{wv}}$, together with the scaling prediction
\begin{equation}
\label{eq:scaling3}
    \frac{L}{L^{(0)}}=\left(\frac{K_\mathrm{wv}}{K_\mathrm{wv}^{(0)}}\right)^{\alpha_\mathrm{cl}/\alpha_\mathrm{wv} }\, .
\end{equation}
Although a positive correlation is observed, it is noticeably weaker than the corresponding correlation with the water--solid adsorption coefficient shown in \Fig~4b of the main text. This difference is consistent with the molecular orientation analysis in Section~\ref{sec:orientation}, which shows that amphiphiles at the contact line lie nearly parallel to the substrate and therefore adopt configurations more similar to those at the water--solid interface than at the water--vapor interface, where they instead orient approximately parallel to the interface~\cite{Staniscia25}.
Consequently, line adsorption is expected to correlate more strongly with $K_{\mathrm{ws}}$ than with $K_{\mathrm{wv}}$.

\section{Continuum model of adsorption at a contact line }
\label{sec:line_model}
\begin{figure}[h]
      \includegraphics[width=0.8\linewidth]{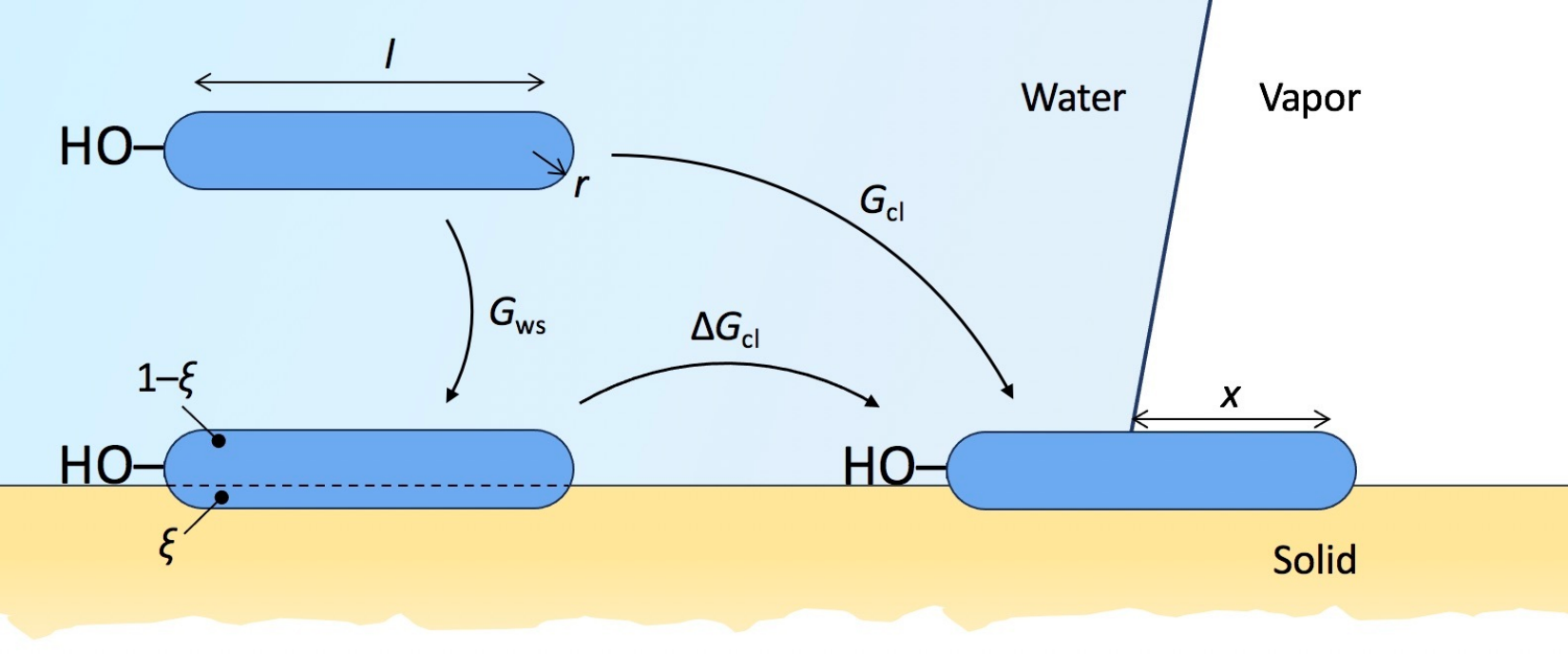}
  \caption{Schematic illustration of the continuum model of amphiphile adsorption at a three-phase contact line. The amphiphilic molecule is modeled as a cylinder of length $l$ and radius $r$. Upon adsorption at the water--substrate interface, a fraction $\xi$ of its surface comes into contact with the substrate, while the remaining fraction $1-\xi$ remains exposed to water. The corresponding adsorption free energy is $G_\mathrm{ws}$. The adsorbed molecule is then transferred across the three-phase contact line, such that a length $x$ protrudes into the vapor phase, with an additional free-energy change $\Delta G_\mathrm{cl}$. The total free energy of transferring the molecule directly from the bulk to the contact line is therefore $G_\mathrm{cl}=G_\mathrm{ws}+\Delta G_\mathrm{cl}$.}
 \label{fig:model}
\end{figure}

We develop a simple continuum model to rationalize the strong affinity of amphiphilic molecules for three-phase contact lines and its dependence on the hydrocarbon chain length. The model estimates adsorption free energies from the changes in interfacial areas when an amphiphile transfers from the bulk water phase to the water--substrate interface and subsequently to the contact line, as illustrated in \Fig~\ref{fig:model}.

\subsection{Adsorption to the water--solid interface}

We approximate the hydrophobic part of an amphiphile as a spherically capped cylinder of radius $r$ and length
$l=a_0 n_\mathrm{C}$, where $n_\mathrm{C}$ is the number of carbon atoms in the hydrocarbon chain and $a_0$ is the molecular length per carbon atom. The polar headgroup ($-$OH) is assumed to remain hydrated, and its contribution to the adsorption free energy is neglected.

When the molecule adsorbs at the water--solid interface, part of its hydrocarbon surface comes into contact with the substrate. We approximate the effective contact area as $A_\mathrm{c}=2\pi r l\xi$, where $\xi$ is the fraction of the cylindrical surface effectively in contact with the substrate. This parameter accounts for the contact geometry, molecular granularity of the hydrocarbon chain and substrate, and possible local substrate deformation. Based on our previous estimates~\cite{Staniscia22a}, we assume $\xi=0.3$--$0.5$.

Formation of the contact replaces water--hydrocarbon and solid--water interfacial areas, each approximated by $A_\mathrm{c}$, with a solid--hydrocarbon interface. The corresponding free-energy change is
\begin{equation}
    G_\mathrm{ws}
    =2\pi r l\xi
    (-\gamma_\mathrm{ow}-\gamma_\mathrm{sw}+\gamma_\mathrm{so}),
    \label{eq:model_gws}
\end{equation}
where $\gamma_\mathrm{ow}$, $\gamma_\mathrm{sw}$, and $\gamma_\mathrm{so}$ denote the oil--water, solid--water, and solid--oil interfacial tensions, respectively. Here, the hydrocarbon part of the amphiphile is represented by an effective oil phase.

Using Young's equations for water and oil,
$\gamma_\mathrm{w}\cos\theta=\gamma_\mathrm{sv}-\gamma_\mathrm{sw}$, and 
$\gamma_\mathrm{o}\cos\theta_\mathrm{o}=\gamma_\mathrm{sv}-\gamma_\mathrm{so}$,
we obtain
\begin{equation}
    G_\mathrm{ws}
    =2\pi r l\xi
    (\gamma_\mathrm{w}\cos\theta
    -\gamma_\mathrm{o}\cos\theta_\mathrm{o}
    -\gamma_\mathrm{ow}).
    \label{eq:Gws}
\end{equation}

We further assume complete wetting of the substrate by the hydrocarbon phase, $\cos\theta_\mathrm{o}=1$, which is a reasonable approximation~\cite{van2002relation}, and apply Antonov's rule, $\gamma_\mathrm{w}\approx\gamma_\mathrm{ow}+\gamma_\mathrm{o}$~\cite{makkonen2018another}. These approximations give
\begin{equation}
    G_\mathrm{ws}
    =-2\pi r l\xi\gamma_\mathrm{w}
    (1-\cos\theta).
    \label{eq:model_gws_final}
\end{equation}

Thus, the adsorption free energy is negative for any nonzero contact angle, indicating favorable adsorption, with its magnitude increasing linearly with hydrocarbon chain length. The corresponding surface adsorption coefficient follows a Boltzmann dependence, $K_\mathrm{ws}=K_\mathrm{ws}^{(0)}\rme^{-\beta G_\mathrm{ws}}$, where $\beta=1/k_\mathrm{B}T$ and the prefactor $K_\mathrm{ws}^{(0)}$ is of the order of a molecular length. Substituting $l=a_0n_\mathrm{C}$, we obtain
\begin{equation}
    K_\mathrm{ws}=K_\mathrm{ws}^{(0)}\rme^{\alpha_\mathrm{ws}n_\mathrm{C}}    
    \label{eq:Kws}
\end{equation}
where we have introduced 
\begin{equation}
    \alpha_\mathrm{ws}
    =2\pi\beta\gamma_\mathrm{w}r a_0\xi
    (1-\cos\theta).
    \label{eq:model_alpha_ws}
\end{equation}

This expression provides a simple physical interpretation of the exponential dependence of surface adsorption on alkyl chain length observed in the simulations. Related interfacial-area arguments were previously developed in Ref.~\citenum{Staniscia22a}.

Using $\gamma_\mathrm{w}=55$~mN/m (SPC/E water), $\theta=120^\circ$, $r=a_0=0.15$~nm, and $T=300$~K, eq~\ref{eq:model_alpha_ws} gives $\alpha_\mathrm{ws}\approx2.83\xi$, or approximately $0.9$--$1.4$ for $\xi=0.3$--$0.5$. The simulation value $\alpha_\mathrm{ws}\approx1.29$ falls in this range.

\subsection{Adsorption to the three-phase contact line}

We next consider transferring an amphiphile from the water--solid interface to the three-phase contact line, as illustrated in \Fig~\ref{fig:model}. Based on the analysis of Section~\ref{sec:orientation}, we assume that the amphiphile remains parallel to the substrate, with its polar headgroup in water and its hydrocarbon chain pointing toward the vapor phase, as observed in the simulations (Section~\ref{sec:orientation}).

Let $x$ denote the length of the cylindrical hydrocarbon segment protruding into the vapor phase ($0\leq x\leq l$). The terminal hydrocarbon group is approximated by a hemispherical cap of radius $r$.

The transfer involves four contributions to the interfacial free energy. First, penetration of the hemispherical cap removes a water--vapor interfacial area of $\pi r^2$. Second, the cap surface (area $2\pi r^2$) becomes exposed to vapor instead of water. 
For simplicity, we neglect the partial overlap of the terminal cap with the substrate and approximate its exposed surface and projected area by those of a full hemisphere.
Third, the fraction $1-\xi$ of the protruding cylindrical surface not in contact with the substrate is transferred from water to vapor. Finally, the displacement exposes additional substrate to water instead of vapor, over an effective area $2\pi rx\xi$.

Adding these contributions, the free-energy difference relative to adsorption at the water--solid interface is
\begin{equation}
    \Delta G_\mathrm{cl}(x)
    =-\pi r^2\gamma_\mathrm{w}+2\pi r^2
    (\gamma_\mathrm{o}-\gamma_\mathrm{ow})
    +2\pi rx(1-\xi)
    (\gamma_\mathrm{o}-\gamma_\mathrm{ow})
    +2\pi rx\xi
    (\gamma_\mathrm{sw}-\gamma_\mathrm{sv}).
    \label{eq:model_delta_g}
\end{equation}
Applying Young's equation for water gives
\begin{equation}
    \Delta G_\mathrm{cl}(x)
    =-\pi r^2
    (\gamma_\mathrm{w}+2\gamma_\mathrm{ow}-2\gamma_\mathrm{o})
    -2\pi rx
    \left[
    (1-\xi)(\gamma_\mathrm{ow}-\gamma_\mathrm{o})
    +\xi\gamma_\mathrm{w}\cos\theta
    \right].
    \label{eq:model_delta_g_simplified}
\end{equation}
Since the free energy depends linearly on $x$, its minimum occurs at either $x=0$ or $x=l$. The fully protruding configuration ($x=l$) is favored when the expression in square brackets in \Eq~\ref{eq:model_delta_g_simplified} is positive. Otherwise, the minimum occurs at $x=0$, with only the terminal cap exposed.

Minimizing the free energy with respect to $x$, $\Delta G_\mathrm{cl}=\min_x\Delta G_\mathrm{cl}(x)$, we obtain
\begin{equation}
    \beta\Delta G_\mathrm{cl}
    =-\alpha_0-\Delta\alpha_\mathrm{cl}n_\mathrm{C},
    \label{eq:model_delta_alpha}
\end{equation}
where
\begin{equation}
    \alpha_0
    =\pi\beta r^2
    (\gamma_\mathrm{w}+2\gamma_\mathrm{ow}-2\gamma_\mathrm{o})
\end{equation}
and
\begin{equation}
    \Delta\alpha_\mathrm{cl}
    =2\pi\beta r a_0
    \max\left\{0,\,
    (1-\xi)(\gamma_\mathrm{ow}-\gamma_\mathrm{o})
    +\xi\gamma_\mathrm{w}\cos\theta
    \right\}.
    \label{eq:model_delta_alpha_cl}
\end{equation}
Thus, the model separates the additional contact-line adsorption free energy into a chain-length-independent contribution from the terminal hydrocarbon group ($\alpha_0$) and a contribution proportional to the chain length ($\Delta\alpha_\mathrm{cl}$). The former favors adsorption at the contact line, whereas the latter provides an additional  free-energy gain after minimization and can further enhance adsorption as the hydrocarbon chain becomes longer.

Taking $\gamma_\mathrm{ow}-\gamma_\mathrm{o}\approx25$ mN/m and $\xi=0.3$--$0.5$, together with the parameters used above, we obtain $\alpha_0\approx1.8$ and $\Delta\alpha_\mathrm{cl}\approx0$--$0.3$. The terminal hydrocarbon group thus provides an additional adsorption free-energy gain of approximately $1.8\,k_\mathrm{B}T$, independent of chain length, while exposure of the hydrocarbon chain can further strengthen adsorption at the contact line. Although sensitive to the effective contact fraction and interfacial tensions, the model provides a simple physical explanation for the enhanced contact-line affinity observed in the simulations.

The total adsorption free energy for transfer from bulk water to the contact line is
\begin{equation}
    G_\mathrm{cl}
    =G_\mathrm{ws}+\Delta G_\mathrm{cl},
\end{equation}
or
\begin{equation}
    \beta G_\mathrm{cl}
    =-\alpha_0
    -(\alpha_\mathrm{ws}+\Delta\alpha_\mathrm{cl})
    n_\mathrm{C}.
\end{equation}

Consequently, the line adsorption coefficient follows
\begin{equation}
    L=L^{(0)}
    \rme^{\alpha_\mathrm{cl}n_\mathrm{C}},
    \label{eq:L}
\end{equation}
with
\begin{equation}
    \alpha_\mathrm{cl}
    =\alpha_\mathrm{ws}+\Delta\alpha_\mathrm{cl}.
    \label{eq:model_alpha_cl}
\end{equation}

Thus, the model predicts an exponential increase of the line adsorption coefficient with hydrocarbon chain length. When protrusion of the hydrocarbon chain is energetically favorable, the additional contribution $\Delta\alpha_\mathrm{cl}>0$ makes this dependence stronger than for adsorption at the water--solid interface. The simulations give $\alpha_\mathrm{cl}\approx1.59$ and $\alpha_\mathrm{ws}\approx1.29$, corresponding to $\Delta\alpha_\mathrm{cl}\approx0.30$, near the upper end of the model estimate ($\Delta\alpha_\mathrm{cl}\approx 0$--$0.3$).

The model therefore offers a physical interpretation of the scaling exponents reported in the main text. In particular, the additional free-energy gain associated with exposing the hydrocarbon chain at the contact line to the vapor phase can explain why $\alpha_\mathrm{cl}>\alpha_\mathrm{ws}$ and why the scaling exponent $\alpha_\mathrm{cl}/\alpha_\mathrm{ws}>1$. Although the model neglects molecular conformational fluctuations, detailed hydration effects, and the microscopic structure of the contact line, it illustrates how molecular geometry and interfacial free energies can govern the relative affinities for the contact line and neighboring interface.

\section{Effect of substrate wettability}

We investigate how line adsorption depends on substrate wettability by simulating droplets with propanol (C3) on substrates with contact angles of $\theta\approx 41^\circ$, $73^\circ$, and $96^\circ$. The wettability was varied by rescaling the partial charges of the substrate headgroups, following our previous work~\cite{Kanduc2017a, Kanduc2017b, Staniscia22a}. This modifies the water--solid interfacial tension and, consequently, the contact angle.

\begin{figure}[h]
\centering
  \includegraphics[width=0.6\linewidth]{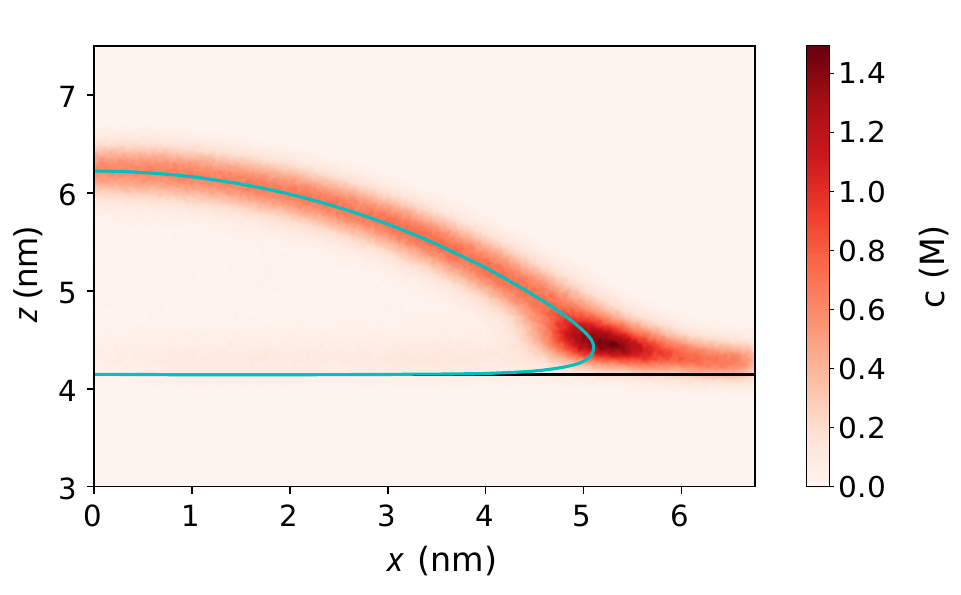}
  \caption{Concentration map of propanol (C3) at a bulk concentration of $c_0\approx23$~mM for a droplet on the most hydrophilic substrate ($\theta\approx41^\circ$).  In addition to adsorption at the three-phase contact line, propanol forms a precursor-like film extending beyond the droplet.}
  \label{fig:hydrophilic}
\end{figure}

For the most hydrophilic substrate ($\theta\approx41^\circ$), propanol also adsorbs onto the solid--vapor interface outside the droplet, forming a precursor-like film of amphiphiles, as shown in \Fig~\ref{fig:hydrophilic}. The presence of this film complicates the separation of surface and line adsorption. A systematic investigation of this regime is beyond the scope of the present work.

\begin{figure*}
     \includegraphics[width=0.35\linewidth]{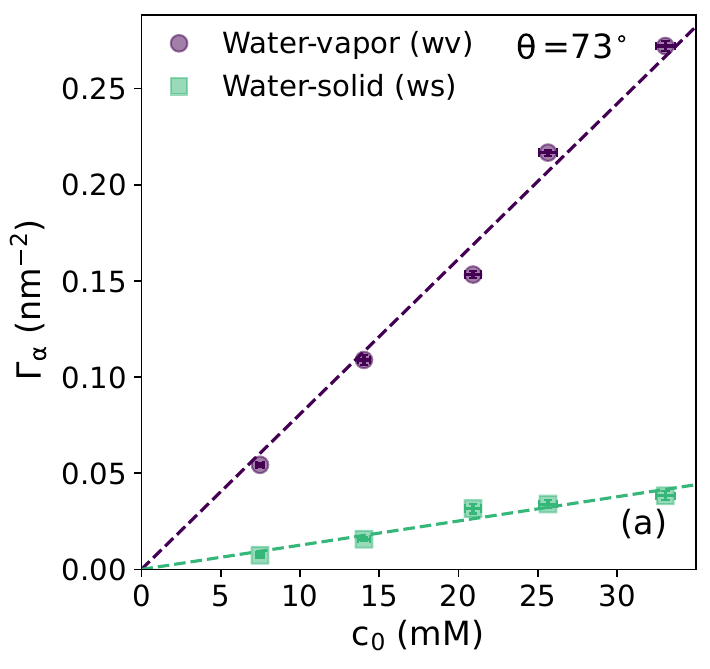}
    \includegraphics[width=0.33\linewidth]{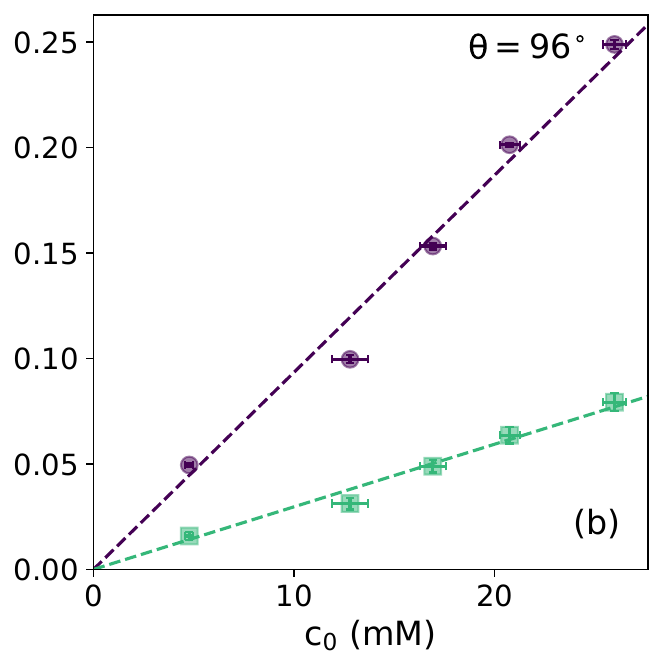}\\
    \includegraphics[width=0.35\linewidth]{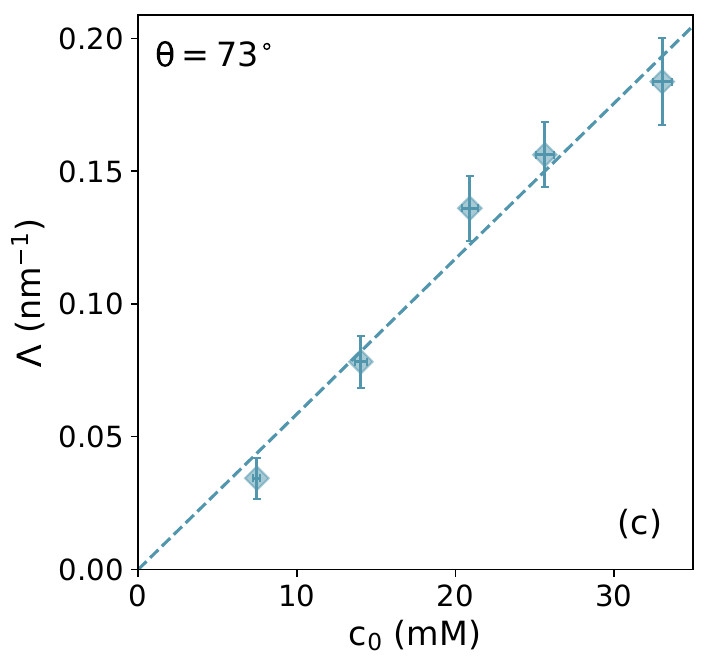}
        \includegraphics[width=0.33\linewidth]{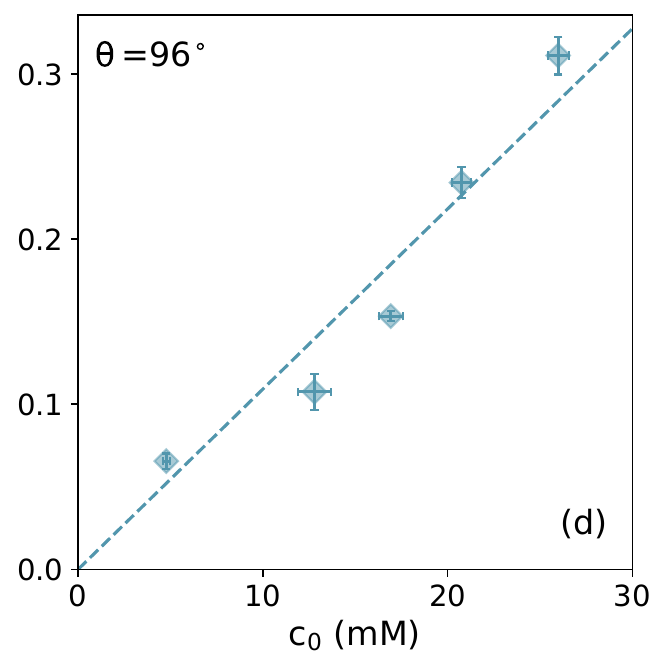}
  \caption{Surface (a,b) and line (c,d) adsorption isotherms of propanol (C3) on substrates with contact angles of $96^\circ$ (left) and $73^\circ$ (right). Dashed lines represent Henry's law fits in the dilute regime. }
  \label{fig:prop_surf_line_ads}
\end{figure*}

For the two less hydrophilic substrates ($\theta\approx73^\circ$ and $96^\circ$), no such film forms, allowing us to apply the same analysis as in the main text. We determined the surface and line adsorption isotherms as functions of the bulk propanol concentration and fitted their dilute regimes with Henry's law to obtain the corresponding adsorption coefficients (\Fig~\ref{fig:prop_surf_line_ads}).

\begin{figure}[h]
\includegraphics[width=0.40\linewidth]{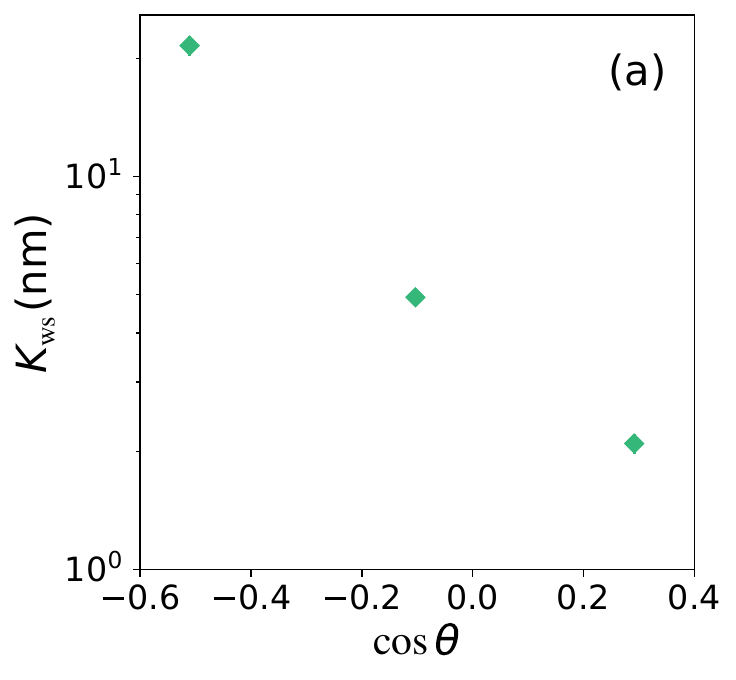}
\includegraphics[width=0.41\linewidth]{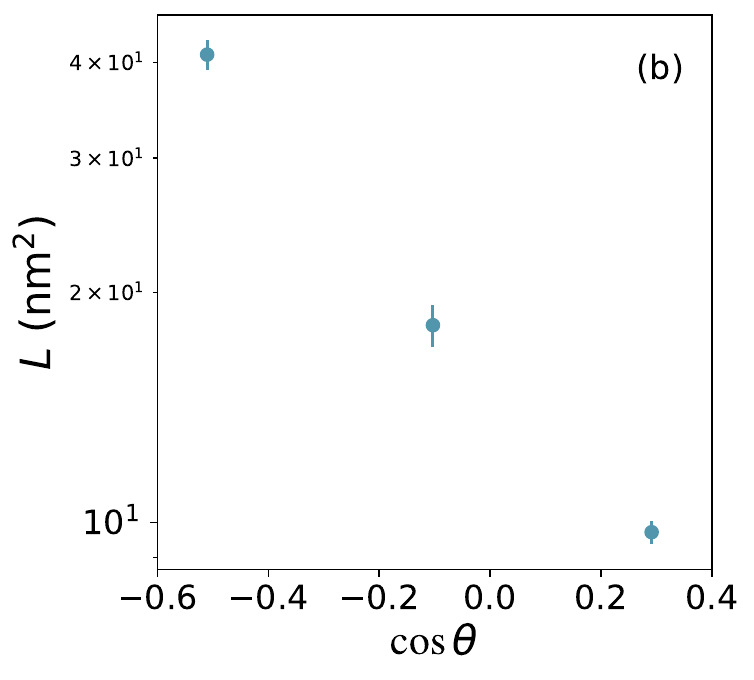}
\caption{(a) Water--solid surface adsorption coefficient and (b) line adsorption coefficient of propanol (C3) as functions of the cosine of the substrate contact angle. }
\label{fig:ads_coeffs_vs_substr}
\end{figure}

The resulting adsorption coefficients are plotted against the cosine of the substrate contact angle in \Fig~\ref{fig:ads_coeffs_vs_substr}. Both the water--solid surface adsorption coefficient and the line adsorption coefficient exhibit an approximately exponential dependence on $\cos\theta$. The former is consistent with previous findings~\cite{Staniscia22a}, while the latter agrees with the continuum model developed in Section~\ref{sec:line_model}.

\section{Line tension reduction}

\begin{figure*}
  \includegraphics[width=0.41\linewidth]{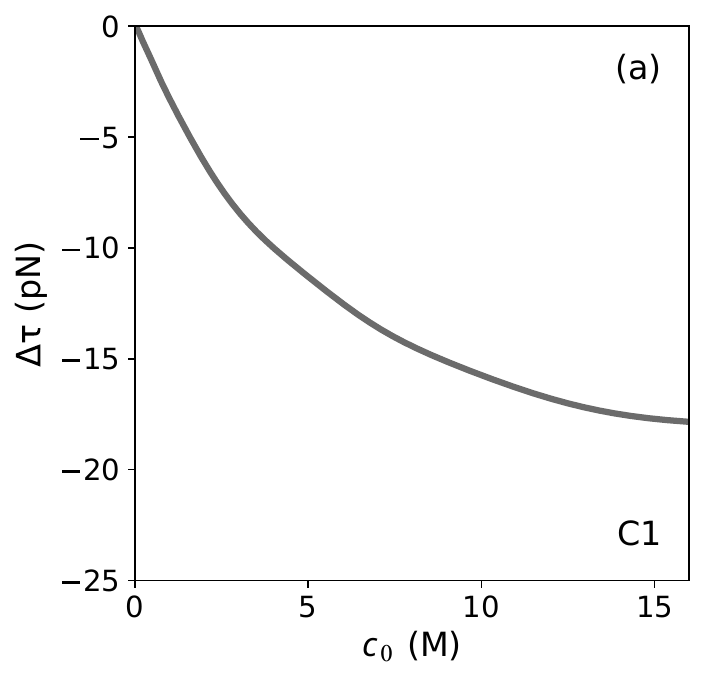}
  \includegraphics[width=0.42\linewidth]{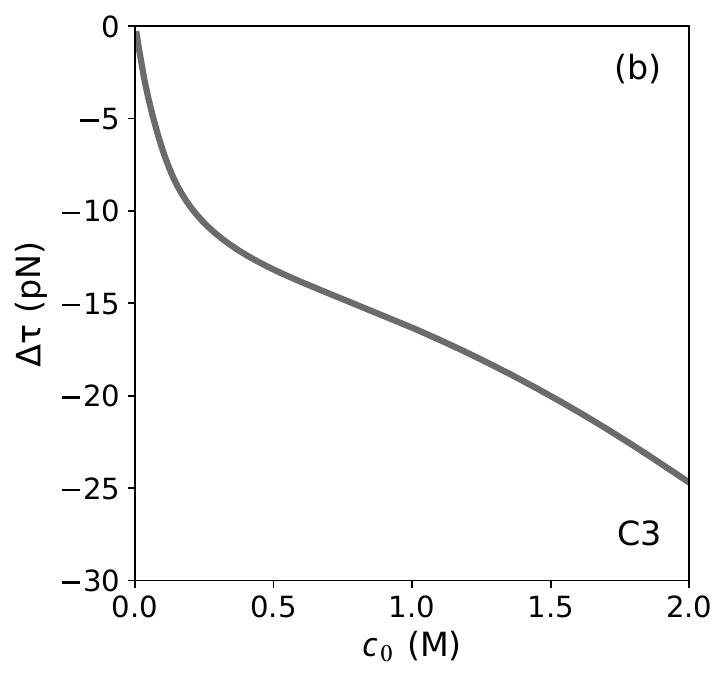}
      \includegraphics[width=0.41\linewidth]{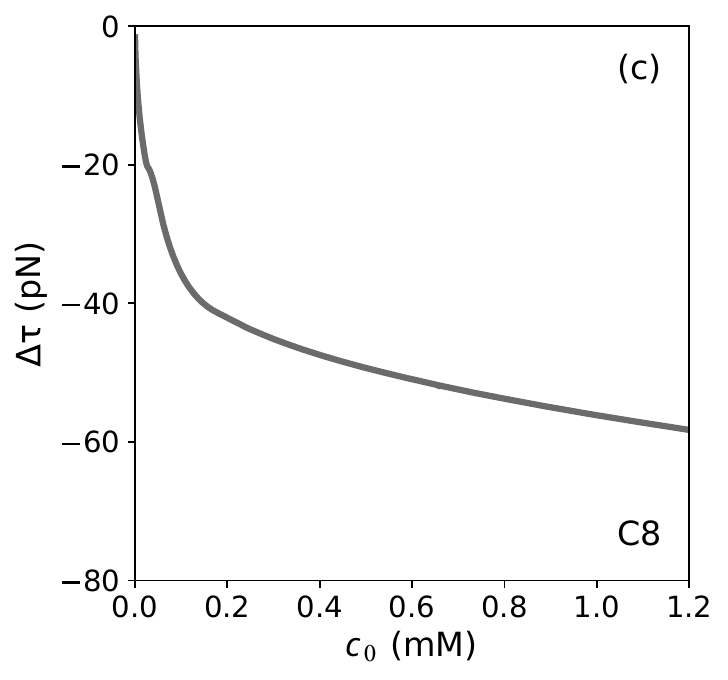}    \includegraphics[width=0.41\linewidth]{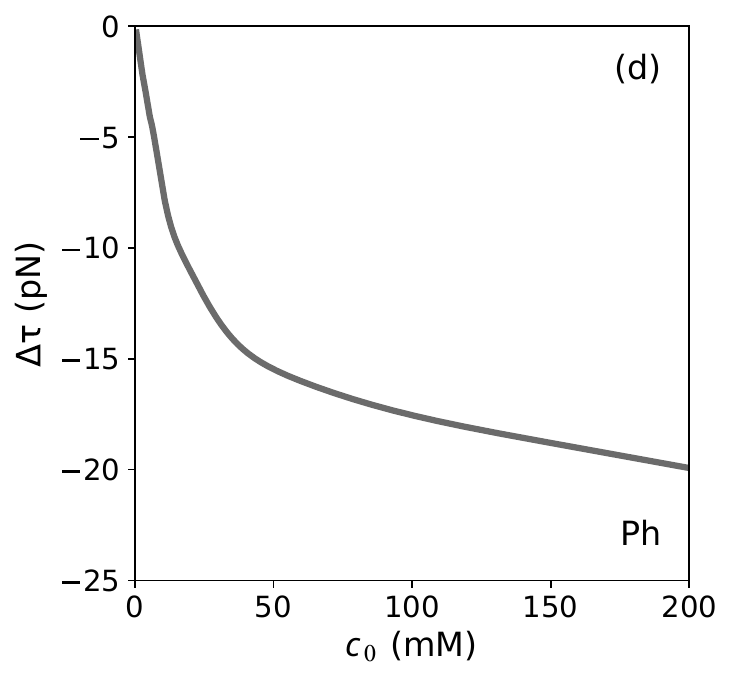}           
  \caption{Reduction in line tension, $\Delta\tau$, as a function of bulk amphiphile concentration for (a) methanol, (b) propanol, (c) octanol, and (d) phenol. The curves were obtained by numerically integrating the Gibbs adsorption relation (\Eq~8 in the main text).}
  \label{fig:detlatau}
\end{figure*}

Similar to the pentanol results shown in \Fig~6 of the main text, \Fig~\ref{fig:detlatau} presents the reduction in line tension due to adsorption of the remaining amphiphiles (except nitrobenzene) investigated in this work.
The curves were obtained by numerically integrating \Eq~8. As for pentanol, the line tension exhibits a characteristic crossover around $c^*$, with the crossover shifting systematically toward lower concentrations as amphiphile size increases.

\section{Effect of line tension on heterogeneous bubble nucleation}

\begin{figure}[h]
    \includegraphics[width=0.5\linewidth]{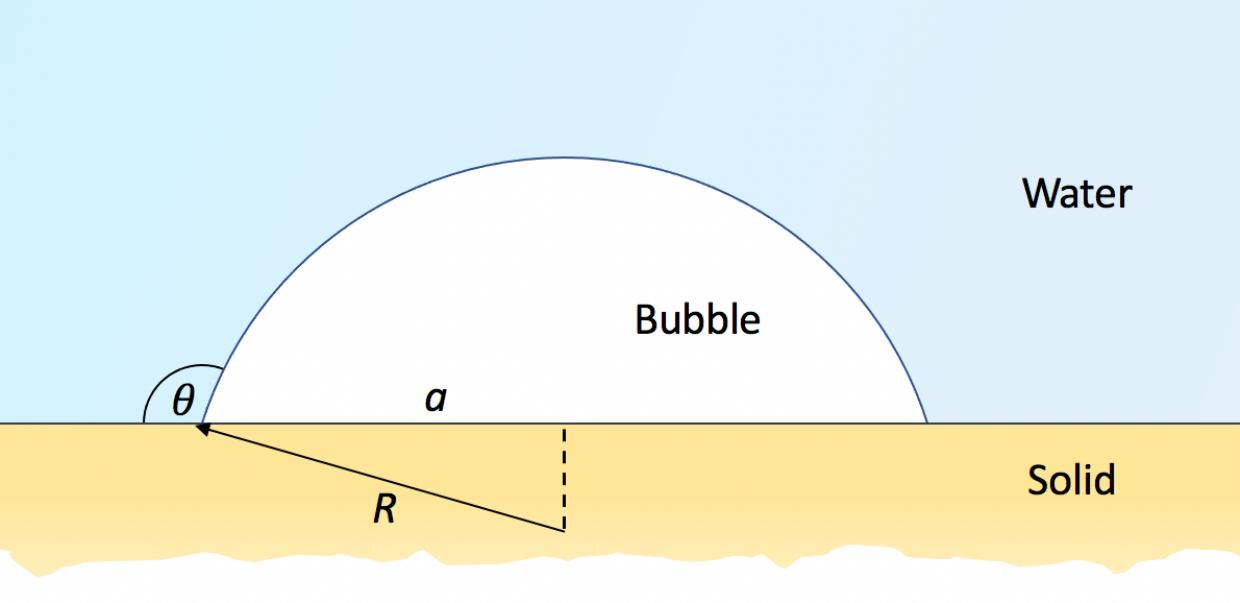}
    \caption{Schematic of a bubble nucleating at a solid surface, with the contact angle $\theta$ measured through the water phase.}
    \label{fig:nucleation}
\end{figure}

We consider heterogeneous nucleation of a bubble at a flat solid surface, as illustrated in \Fig~\ref{fig:nucleation}, and examine how changes in line tension affect the nucleation threshold. We first consider cavitation in a liquid under negative pressure and then reformulate the analysis for bubble nucleation in a solution supersaturated with a noncondensable gas.

\subsection{Cavitation under negative pressure}

Within classical nucleation theory (CNT), the free energy barrier for the formation of a spherical-cap vapor bubble on a solid surface in a liquid under negative pressure is~\cite{caupin2006cavitation}
\begin{equation}
G^*=\frac{16\pi\gamma^3}{3p^2}h(\theta),
\end{equation}
where $\gamma$ is the liquid--vapor surface tension, $p<0$ is the ambient liquid pressure, and $\theta$ is the contact angle measured through the liquid. The geometrical factor is~\cite{caupin2006cavitation}
\begin{equation}
h(\theta)=\frac{2+3\cos\theta-\cos^3\theta}{4}.
\end{equation}
This expression neglects line tension. Line tension contributes an additional free energy term of $2\pi a\tau$, where $a$ is the base radius of the bubble.
Treating the line-tension contribution as a small perturbation, the corresponding first-order change in the nucleation barrier at a given pressure is 
\begin{equation}
\Delta G_{\mathrm{line}}^*
=2\pi R^*\sin\theta\,\tau
=-\frac{4\pi\gamma\sin\theta}{p}\tau,
\end{equation}
where $R^*=-2\gamma/p$ is the critical radius of the unperturbed bubble.

We now estimate how a finite line tension changes the cavitation pressure. 
The cavitation pressure $p_\mathrm{cav}$ is defined as the pressure at which the cavitation probability becomes appreciable on a given observation length- and timescale~\cite{caupin2006cavitation}. Assuming that the cavitation onset occurs at a specific nucleation barrier, the shifted cavitation pressure $p_\mathrm{cav}'=p_\mathrm{cav}+\Delta p_\mathrm{cav}$ satisfies
\begin{equation}
\frac{16\pi\gamma^3 h(\theta)}{3(p_\mathrm{cav}+\Delta p_\mathrm{cav})^2}
-\frac{4\pi\gamma\sin\theta}{p_\mathrm{cav}+\Delta p_\mathrm{cav}}\tau
=
\frac{16\pi\gamma^3 h(\theta)}{3p_\mathrm{cav}^2}.
\end{equation}
Expanding to first order in $\Delta p_\mathrm{cav}$ and $\tau$ yields
\begin{equation}
\frac{\Delta p_\mathrm{cav}}{p_\mathrm{cav}}
=-\frac{3p_\mathrm{cav}\sin\theta}
{8\gamma^2h(\theta)}\tau
\label{eq:dpcav}
\end{equation}
Thus, the cavitation pressure depends linearly on the line tension $\tau$. Consequently, a reduction in line tension ($\Delta\tau<0$), allowing bubble nucleation at a less negative liquid pressure.

As an example, we consider a smooth hydrophobic surface with a contact angle of $\theta=120^\circ$, for which CNT combined with molecular dynamics simulations predicts a cavitation pressure of $p_\mathrm{cav}\approx-60$ MPa~\cite{sako2025nanoscale}. Taking $\gamma=72$ mN/m, \Eq~\ref{eq:dpcav} predicts that line-tension reductions of $\Delta\tau=-10$ and $-20$ pN increase the cavitation pressure (i.e., decrease its magnitude) by approximately 14 and 29 MPa, respectively. These shifts correspond to reductions in the magnitude of the cavitation pressure by approximately 24\% and 48\%, respectively. Although such large shifts lie beyond the validity of the perturbative approximation, they illustrate the potentially substantial effect of line adsorption on heterogeneous cavitation.

\subsection{Bubble nucleation from supersaturated solutions}

The same approach can be applied to heterogeneous nucleation of bubbles containing a noncondensable gas in a supersaturated aqueous solution. In this case, the nucleation driving pressure is $p_\mathrm{g}^\mathrm{eq}-p$, where $p_\mathrm{g}^\mathrm{eq}$ is the gas pressure in equilibrium with the dissolved gas and $p$ is the ambient liquid pressure~\cite{jones1999bubble}. Assuming Henry's law, we have $p_\mathrm{g}^\mathrm{eq}=Sp$, where $S$ is the gas supersaturation relative to equilibrium at the ambient pressure $p$. 

We define the critical supersaturation $S_\mathrm{c}$ as the supersaturation at which the nucleation probability becomes appreciable on a given observation length- and timescale.
Replacing $p_\mathrm{cav}$ in \Eq~\ref{eq:dpcav} by $p(1-S_\mathrm{c})$ and assuming $S_\mathrm{c}\gg1$, we obtain
\begin{equation}
\frac{\Delta S_\mathrm{c}}{S_\mathrm{c}}
\simeq
\frac{3S_\mathrm{c}p\sin\theta}
{8\gamma^2h(\theta)}
\tau
\label{eq:dsnucl}
\end{equation}
Thus, a reduction in line tension lowers the critical supersaturation required for bubble nucleation, with the relative effect increasing with supersaturation.

High gas supersaturations can occur during electrochemical gas generation in water. For example, Chen et al.~\cite{chen2014electrochemical} and German et al.~\cite{german2016electrochemistry} reported critical H$_2$ supersaturations of approximately $S_\mathrm{c}=300$ at Pt nanoelectrodes. Taking $S_\mathrm{c}=300$, $p=1$ atm, $\theta=120^\circ$, and $\gamma=72$ mN/m, a line-tension reduction of $\Delta\tau=-20$ pN reduces the critical supersaturation by approximately 25\%, according to our first-order estimate. Both studies also observed that surfactants facilitate bubble nucleation by lowering the critical supersaturation, although their interpretation focused on changes in surface tension without considering line tension.

\section{Calculation of adsorption excesses}

Surface and line adsorption are calculated from the two-dimensional density maps of water and amphiphiles. Because a droplet contains bulk, interfacial, and contact-line regions with distinct physical properties, the first step is to identify these regions and the corresponding Gibbs dividing surfaces. The adsorption quantities are then obtained from the resulting geometrical decomposition.

\subsection{Construction of density maps}

\begin{figure}[h]
      \includegraphics[width=0.8\linewidth]{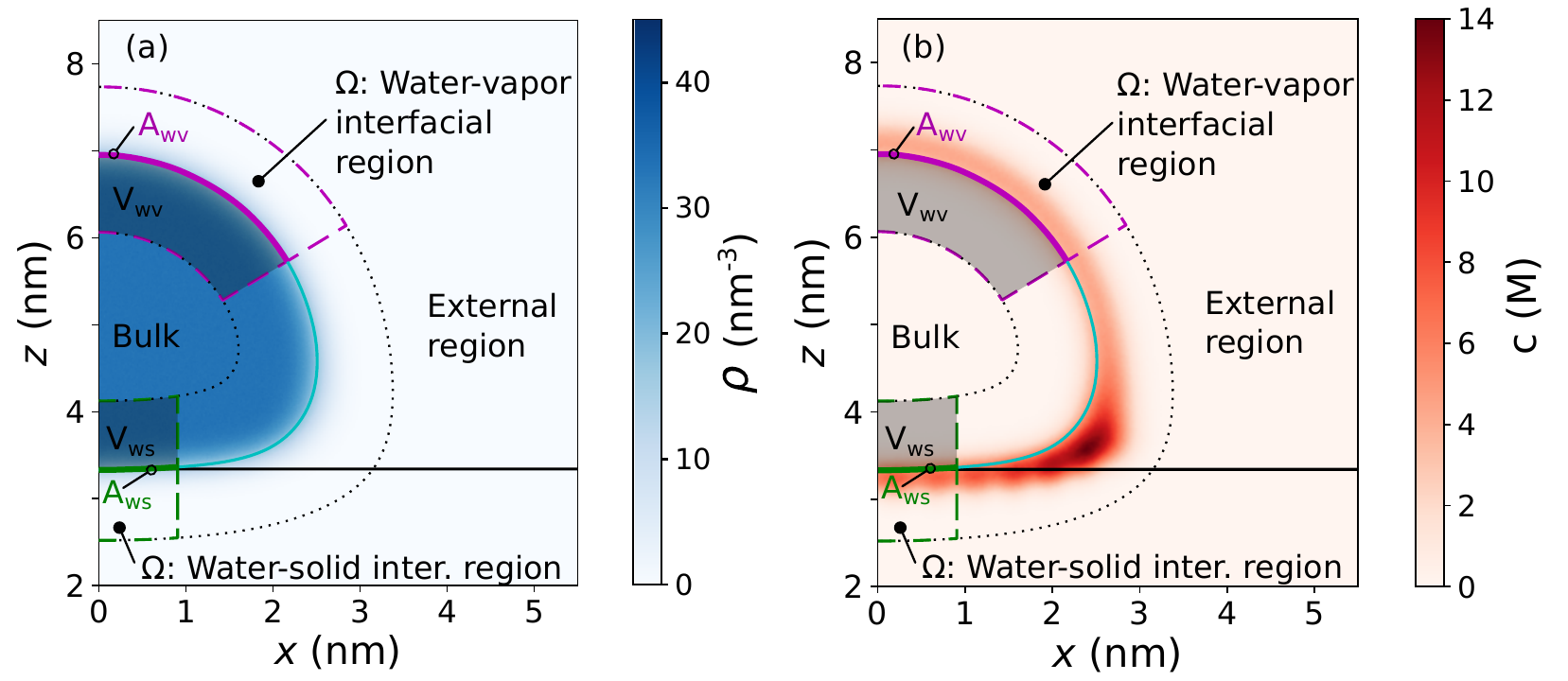}
  \caption{Two-dimensional (a) water density map and (b) pentanol concentration map at $c_0\simeq5$~mM.  Dotted lines delineate the bulk, interfacial, and external regions. Dashed purple and green curves mark the water--vapor and water--solid regions, respectively. The corresponding GDS are shown as solid curves of the same colors, while the cyan curve denotes the GDS constructed from the entire interfacial region. The shaded regions represent the volumes $V_\alpha$ and the corresponding colored curves the interfacial areas $A_\alpha$ used in \Eqs~\ref{eq:excesss_dens} and~\ref{eq:excesss_dens_c}. The solid black line indicates the substrate. }
 \label{fig:comparts}
\end{figure}

The first step is to construct two-dimensional concentration (i.e., number density) maps of water and amphiphiles from the trajectories, as shown in \Fig~\ref{fig:comparts}. Molecular center-of-mass positions are expressed relative to the droplet center of mass (computed from the water molecules) in each frame and averaged over simulation time and independent simulation replicas.

\subsection{Partitioning of the density maps}
The second step is to partition the density maps into four regions corresponding to distinct physical environments: (i) the bulk region, where the water and amphiphile densities are uniform; (ii) the external vapor region, where both densities are negligible; and the (iii) water--vapor and (iv) water--solid interfacial regions, where the densities (concentrations) vary continuously between the adjacent phases.

The boundaries of these regions are identified from the density gradients. The numerical algorithm, implemented in Python, is available in the repository and is briefly summarized here.

First, the density maps are smoothed by convolution with a two-dimensional Gaussian kernel to reduce statistical noise. Starting from an initial iso-density contour, the local gradients of both density fields are evaluated, and the contour is iteratively displaced toward either the bulk or the external region until both gradients fall below prescribed thresholds. The boundary is thus determined by whichever density field reaches its threshold last. The average gradient magnitudes along the resulting boundaries are typically $\sim 10^{-3}$~nm$^{-4}$ for the bulk boundary and $\sim 10^{-4}$~nm$^{-4}$  for the external boundary.

The amphiphile concentration maps in Fig. \ref{fig:comparts}b show approximately uniform adsorption along the water--vapor and water--solid interfaces sufficiently far from the contact line. The corresponding interfacial regions are therefore defined by excluding the vicinity of the contact line, where the two interfaces overlap. Their boundaries toward the contact-line region are chosen perpendicular to the respective interfaces. The positions of these boundaries are manually varied to identify regions where the calculated surface excesses,  $\Gamma_\mathrm{ws}$ and $\Gamma_\mathrm{wv}$, remain approximately constant.

\subsection{Gibbs dividing surfaces}

The next step is to identify the equimolar Gibbs dividing surfaces (GDS) within 
water--vapor ($\alpha=\mathrm{wv}$) and water--solid ($\alpha=\mathrm{ws}$) regions. For the  water--solid interface within the  water--solid region, the GDS is conventionally defined as the surface for which the excess water density vanishes, $\Gamma^{(\mathrm{w})}=0$. Here we generalize this definition to curved interfaces, which we need for the water--vapor interface, by expressing the excess water density as a function of the iso-density value $\rho^{\prime}$ defining a candidate interface,
\begin{equation}\label{eq:excesss_dens}
    \Gamma^{(\mathrm{w})}_{\alpha} ( \rho^{\prime} ) \equiv  \frac{1}{A_{\alpha}(\rho^{\prime})}\, \int_\Omega \mbox{d}\mathbf{r} \, \left[ \rho(\mathbf{r})-\rho_\mathrm{v}\right] -  \frac{V_{\alpha}(\rho^{\prime})}{A_{\alpha}(\rho^{\prime})} \, (\rho_0-\rho_\mathrm{v})\, ,
\end{equation}
where $\rho(\mathbf{r})$ is the water density, $\rho_0$ and $\rho_\mathrm{v}$ are the bulk and vapor water densities, respectively, and $\Omega$ denotes either the water--vapor or the water--solid region. Here, $V_{\alpha}(\rho^{\prime})$ and $A_{\alpha}(\rho^{\prime})$ are the corresponding water volume and interfacial area within the region $\Omega$, highlighted in \Fig~\ref{fig:comparts}.
The GDS for a corresponding region is then defined by the iso-density value $\rho_{\alpha}^{\mathrm{GDS}}$ satisfying
$\Gamma^{(\mathrm{w})}_{\alpha}(\rho_{\alpha}^{\mathrm{GDS}})=0$.

Separate GDSs are constructed for the water--vapor and water--solid interfaces because adsorption is evaluated independently at each interface. The resulting GDSs are shown in Figure~7 of the main text and in Figure~\ref{fig:comparts}.

For completeness, we also applied the same procedure to the entire interfacial region, thereby constructing a single GDS for the complete droplet. For the systems considered here, this interface differs only negligibly from the combination of the separately constructed water--vapor and water--solid GDSs, and is used in this work only to visualize the droplet shape.

\subsection{Calculation of amphiphile adsorption}

The amphiphile adsorption is calculated using the same excess-density formalism as employed for the Gibbs dividing surfaces. In \Eq~\ref{eq:excesss_dens}, the water density $\rho(\mathbf{r})$ is replaced by the amphiphile concentration $c(\mathbf{r})$, the bulk water density by the bulk amphiphile concentration $c_0$, and the amphiphile concentration in the external region is neglected. This yields
\begin{equation}\label{eq:excesss_dens_c}
    \Gamma_{\alpha} =  \frac{1}{A_{\alpha}(\rho_{\alpha}^{\mathrm{GDS}})}\, \int_\Omega \mbox{d}\mathbf{r} \, c(\mathbf{r}) -  \frac{V_{\alpha}(\rho_{\alpha}^{\mathrm{GDS}})\,c_0}{A_{\alpha}(\rho_{\alpha}^{\mathrm{GDS}})}\, .
\end{equation}
Evaluating this expression over the water--vapor and water--solid interfacial regions gives the corresponding surface excesses, $\Gamma_{\mathrm{wv}}$ and $\Gamma_{\mathrm{ws}}$.

Finally, the surface excesses $\Gamma_{\mathrm{wv}}$ and $\Gamma_{\mathrm{ws}}$, together with the total number of amphiphiles $N$ and the geometrical properties of the droplet obtained from the Gibbs dividing surfaces, are substituted into \Eq~2 of the main text to determine the line adsorption, $\Lambda$.

\setlength{\bibsep}{0pt}
\bibliography{bibliography}

%\fi